\documentclass{IEEEtran}
\usepackage{cite}
\usepackage{amsmath,amssymb,amsfonts}
\usepackage{algorithmic}
\usepackage{graphicx}
\usepackage{subcaption}
\usepackage{dsfont}
\usepackage{url}
\usepackage[table]{xcolor}
\usepackage{multicol,longtable}
\usepackage{textcomp}

\newtheorem{definition}{Definition}[section]

\newtheorem{proposition}{Proposition}[section]

\begin{document}

\title{Hybrid modeling: Applications in real-time diagnosis}
\author{{Ion Matei},
{Johan de Kleer,  Alexander Feldman, Rahul Rai and Souma Chowdhury}}

\maketitle

\begin{abstract}
Reduced-order models that accurately abstract high fidelity models and enable faster simulation is vital for real-time, model-based diagnosis applications. In this paper, we outline a novel hybrid modeling approach that combines machine learning inspired models and physics-based models to generate reduced-order models from high fidelity models. We are using such models for real-time diagnosis applications. Specifically, we have developed machine learning inspired representations to generate reduced order component models that preserve, in part, the physical interpretation of the original high fidelity component models. To ensure the accuracy, scalability and numerical stability of the learning algorithms when training the reduced-order models we use optimization platforms featuring automatic differentiation. Training data is generated by simulating the high-fidelity model. We showcase our approach in the context of fault diagnosis of a rail switch system. Three new model abstractions whose complexities are two orders of magnitude smaller than the complexity of the high fidelity model, both in the number of equations and simulation time are shown. The numerical experiments and results demonstrate the efficacy of the proposed hybrid modeling approach.
\end{abstract}

\section{Introduction}
\label{sec:introduction}
Model-based fault diagnosis comprises determining a fault by comparing deviations (of at least one characteristics property or parameter of the system) from nominal conditions in system observable quantities. The diagnosis engine typically is provided with a model of the system, nominal values of the
parameters of the model, and a number of  inputs
and outputs values of the system. The main goal of a diagnosis engine is to determine the kind, size, location, and detection time of a fault and to isolate it. At the core of any diagnosis engine are models, which can be typically described using three classes: \textit{black box} (data-driven), \textit{white box} (physics-based) and \textit{grey box} (hybrid). Black box models embed the input-output behavior of a physical system, without any physical interpretation of how the data is processed inside the model. Machine learning and system identification models belong to the black box model category. White box models models are constructed from first principles. For complex systems, it is non-trivial to build high quality first-principle models. They require expert knowledge and often proprietary information that is not always available. Even when built, they may be too complicated to be of use in real-time applications. Grey box models are \textit{hybrid} models representing a  combination of physics-based models and data-driven models.

There are two main diagnosis modeling approaches: (1) \textit{model-based diagnosis} and (2) \textit{data-driven diagnosis}.
There is a long history of model-based diagnosis methods proposed independently by the artificial intelligence \cite{deKleer92,DEKLEER200325,de1987diagnosing,WILLIAMS19911}, and control \cite{Gertler_1998,Isermann200571,Isermann1997639,Patton_2000} communities.
These methods can be further classified into three major categories: 1)  parameter estimation based, 2) observers-based, and 3) redundancy relations (or parity equations)-based. In the case of parameter estimation based diagnosis \cite{ISERMANN1993815,ISERMANN_2007}, we track and detect changes in the  system parameters (e.g., stiffness or
damping coefficients). If a significant deviation is detected an alarm is raised. The parameter tracking can be performed by running optimization algorithms using time series of sensor measurements that learn the current parameter values, or filters (e.g., Kalman filter \cite{Kalman}, particle filter \cite{Arulampalam02atutorial}), where the state vector is augmented to include system parameters.  If the nominal values of the system parameters are known, state and output observers can be used, where deviations in the system behavior are treated as additive faults. There are several possible combinations of observers aimed at detecting single or multiple faults:  observer excited by one output \cite{Clark1978ASI}, bank of observers excited by all outputs \cite{WILLSKY1976601}, bank of observers excited by single outputs \cite{Clark1978ASI}, or bank of observers excited by all outputs, except one \cite{FRANK198763}. The redundant relations (or parity equations) diagnosis methods use part of the system model (constitutive equations) to derive residuals \cite{STAROSWIECKI2000301,STAROSWIECKI2001687}. Faults are detected when the residual values are larger than some preset threshold. By tracking the residual sensitivity to system parameters or components, faults can be isolated as well. The parameter estimation and observer based diagnosis methods require a fault model: deviation of parameters from their nominal value, additive or multiplicative faults. The redundancy relations based methods do not require fault models, however they typically require more sensors to disambiguate between multiple possible faults.

Model-based diagnosis requires accurate models to detect and isolate faults in physical systems. For real-time diagnosis, such models need to simulate within a maximum allotted time interval. Typically, the more accurate models are, the more complex they become and hence it takes more time to simulate them. In addition, the type of mathematical model, e.g., ordinary differential equation (ODE), differential algebraic equation (DAE), and its dynamic response, e.g., stiff, or non-stiff, can add to the time complexity of the simulation process. In particular, DAE simulations require the use of the Newton-Rhapson algorithm that in turn requires the inversion of the system Jacobian matrix. Such an operation has $O(n^3)$ time complexity, where $n$ is the number of equations.

Data driven diagnosis methods have the advantage of not having to simulate system models for decision making, but they require large training data sets covering the faulty behavior. One approach to cope with insufficient training data is to generate synthetic data for faults for which no data exist \cite{Minhas2014,Saha2014}. Data-driven models (e.g., classifiers based on neural networks, random forests, or decision trees) extract structure ignored in hand crafted models, but they are not reliable for corner cases. Nonetheless, classifiers trained on data represented in the frequency domain proved to be useful in rotating machinery diagnosis \cite{Vania2013,Ishibashi2017,7983338}, and in particular for bearing diagnosis \cite{BRKOVIC201763,Hoang2017,7836314}.

Hybrid models combine the best of both worlds. They have the advantage of preserving, in part, the physical interpretation of the system model, which is particularly useful in diagnosis and prognosis applications. From the learning perspective, using the partial system information leads to a decrease in the data-driven model complexity and an increase in the generalizability\footnote{By generalizability we mean the prediction accuracy on data not used for training.} of the hybrid model. The reduced complexity of the data-driven model is based on the observation that we only need to incorporate the missing behavior which is less complex than the overall behavior. The increase in generalizability comes from the fact that the partial physics-based model induces an inherent regularization function that constrains the data-driven model parameter search space.

In this paper we propose a hybrid modeling approach to reduce the complexity of physic-based models and enable real time diagnosis based on parameter estimation. We showcase our approach in the case of a rail switch system. In particular, the main contributions of the paper are as follows:
\begin{itemize}
    \item We design a hybridization technique that combines physics based models with data-driven inspired models while preserving composability.
    \item We introduce three representations of data-driven inspired models that preserve physical interpretability and numerical stability.
    \item We show how automatic differentiation can be used to train hybrid models in synergy with ODE and DAE solvers.
    \item We demonstrate how the hybrid modeling approach is used to reduce the complexity of a rail switch model by two orders of magnitude.
    \item We demonstrate how the reduce complexity rail switch model is used for fault diagnosis based on tracking system parameters.
\end{itemize}

\textbf{Paper structure}: Section \ref{sec_problem_description} describes the main steps of the hybrid modeling approach, the model assumptions and the diagnosis approach. In Section \ref{sec_rail_switch_model} we introduce the rail switch system and its corresponding physical model, whereas in Section \ref{sec_fault_augmentation} we present the list of faults and their parameterization. Section \ref{sec_acausal_modeling} describes the constructs and representations used for hybrid modeling. In Section \ref{sec_hybrid_rail_switch_model} we demonstrate the hybrid modeling approach on the rail switch model and show the reduction in model complexity while preserving accuracy. Finally, Section \ref{sec_fault_diagnosis} shows the diagnosis results for the rail switch system when using the hybrid model. A comparison of our hybrid modeling approach with the state of the art is introduced in Section \ref{sec_state_of_the_art}. We end the paper with some conclusions.

\section{Problem description}
\label{sec_problem_description}
\begin{figure}[ht!]
\begin{center}
\includegraphics[scale = 0.2]{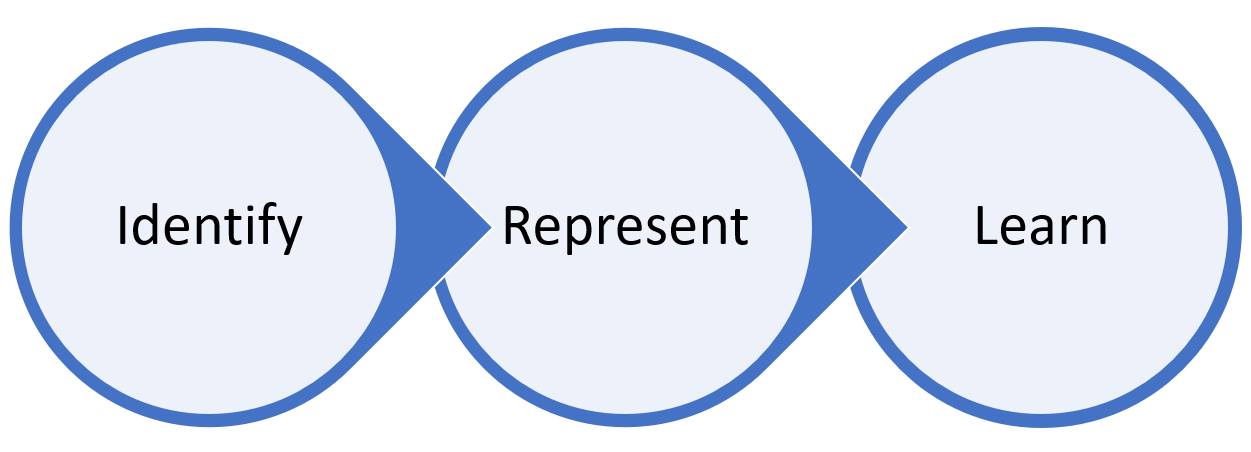}
\end{center}
\caption{Three steps hybrid modeling approach}
\label{fig:02121534}
\end{figure}

The proposed hybrid modeling approach follows three main steps (Figure \ref{fig:02121534}):  (i) \textit{identifying} the system components responsible for making the simulation time large, (ii) finding simpler, parameterized \textit{representation}s for such components, and (iii) \textit{learning} the parameters of the new component represenations
The proposed approach ensures that the chosen representations preserve, in part, the physical meaning of the original physical components. By in part we mean that the variables of the component interface through, which it interacts with the rest of the system components, preserve their physical interpretation.  Such a property is particularly useful in diagnosis since it points to a physical explanation of a faulty behavior.

We consider physical systems whose behavior can be described by a DAE of the form
\begin{eqnarray}
 \label{equ:11132127}
  0 &=& F(\dot{{x}},x,u), \\
 \label{equ:11132128}
  {y} &=& h({x},{u}),
\end{eqnarray}
where ${x}$ represents the state vector, ${u}$ is a vector of inputs, and ${y}$ is a vector of outputs. We consider \textit{parametric faults}: faults that can be described through changes in system parameter values. Parametric faults do not impose  significant constraints on the type of faults that we can detect and isolate. Indeed, as shown in our previous work \cite{Honda2014,Minhas2014,Saha2014}, we can augment the physical model with fault modes inspired by the physics of failure. The physics-based fault augmentation process adds additional equations to the model. These new equations are dependent on parameters whose activation induces the simulated faulty behavior. The type of faults introduced are domain dependent. We can model faults in the electrical (short, open connections, parameter drifts), mechanical (broken flanges, stuck flanges, torque losses due to added friction, efficiency losses), or fluid (blocked pipes, leaking pipes) domains.

Let $\mathcal{F}=\{F_0,F_1, \ldots, F_{L}\}$ denote the set of faults we would like to detect and isolate, where $F_0$ denotes the normal behavior.  The diagnosis objective is to construct a classifier $f: Y \rightarrow \{F_0,F_1, \ldots, F_{L}\}$, where $Y$ is a set of observations of the system behavior, typically given by a set of sensor measurements that are processed sample by sample (online) or as a batch (offline). We associate a set of fault parameters $\{\theta_1,\ldots, \theta_L\}$  to each of the fault mode with nominal values $\theta_i^*$ for $i\in\{1,\ldots,L\}$. The classifier fault detection scheme is defined as a deviation of the output measurements from their expected values. The fault isolation is based on the deviation of the fault parameters from their nominal values, i.e., $\|\theta_i-\theta_i^*\|\geq \epsilon_i$, where $\epsilon_i$ is a fault specific threshold that depends on measurement noise statistics. Several fault parameter deviations are simultaneously possible.  Hence, there may be some ambiguity in the fault diagnosis. Ambiguous fault diagnosis happens when the sensor measurements do not contain enough information to differentiate between distinct faults. 

To better motivate the utility of the proposed hybrid modeling concept and technical challenges involved in the class of diagnosis problems we are interested in, consider the case of developing a predictive model for a cyber-physical system such as a railway switch system (Figure \ref{fig:12121446}). The switch is a microcosm of the modeling  challenges many complex systems have (e.g., nonlinearities, distributed parameters). Rail switch system is an illustrative example and will be our \textit{running example.} Railway switching systems generate approximately 60\% of the failure statistics related to railway traffic disruptions due to signaling problems and thus are vital to monitor through diagnostic approaches. The next section introduces the rail switch system and model.

\section{Rail switch model}
\label{sec_rail_switch_model}

\begin{figure}[ht!]
\begin{center}
\includegraphics[width=0.4\textwidth]{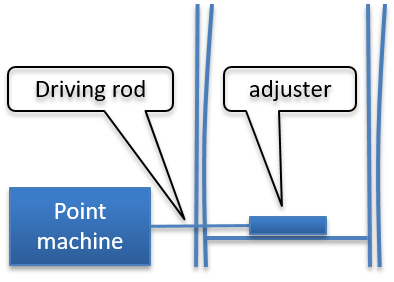}
\end{center}
\caption{Rail-switch schematics.}
\label{fig:12121446}
\end{figure}

\begin{figure}[ht!]
\begin{center}
\includegraphics[width=0.48\textwidth]{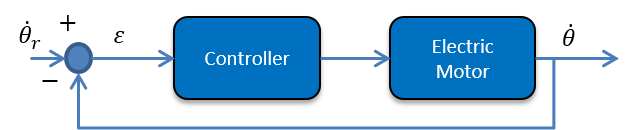}
\end{center}
\caption{Servo-motor schematics.}
\label{fig:03201520}
\end{figure}

\begin{figure}[ht!]
\begin{center}
\includegraphics[width=0.48\textwidth]{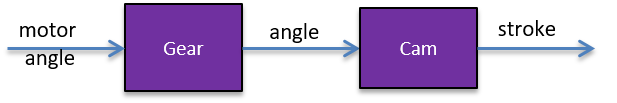}
\end{center}
\caption{Gear mechanism.}
\label{fig:03201534}
\end{figure}

As a case study, we consider a rail switch system used for guiding  trains from one track to another. The schematics of the system is presented in Figure \ref{fig:12121446}, depicting the main components of the rail switch. The rail switch is composed of a point machine and a gear-mechanism. The rail load is composed of a mechanical adjuster and tongue-rails.
The point machine is the component of the rail switch system that is responsible for moving the rails and locking them in their final position until a new motion action is initiated. It is composed of two sub-components: a servo-motor and a gear mechanism. The electric motor transforms electrical energy into mechanical energy and generates a rotational motion. The gear mechanism scales down the angular velocity of the motor and amplifies the torque generated by the motor. In addition, using a cam system, the rotational motion is transformed into translational motion. The servo-motor is composed of two sub-components: an electric motor and a controller (Figure \ref{fig:03201520}). The controller makes sure the motor's angular velocity follows a prescribed reference. The angular velocity is perturbed by the rail load torque. The gear mechanism (Figure \ref{fig:03201534}) is responsible for scaling down the angular velocity generated by the servo-motor, amplifying the torque generated by the servo-motor and transforming the rotational motion into a translational motion.

\begin{figure}[ht!]
\begin{center}
\includegraphics[width=0.48\textwidth]{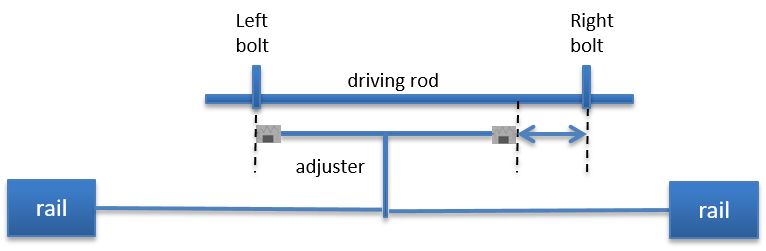}
\end{center}
\caption{Adjuster mechanics.}
\label{fig:03201536}
\end{figure}

The rail load is comprised of an adjuster and  tongue rails. The adjuster``floats'' on the driving rod, and acts as an interface between the driving rod and the rail and transfers the force generated by the motor (trough the driving rod) to the rails, generating this way motion. The adjuster (Figure \ref{fig:03201536}) connects the driving rod attached to the point machine to the rails, and hence it is responsible for transferring the translational motion. There is a delay between the time instances the driving rod and the adjuster start moving. This delay is controlled by two bolts on the driving rod. Tighter bolt settings means a smaller delay, while looser bolt settings produce a larger delay. The adjuster is connected to two rails that are moved from left to right or right to left, depending on the traffic needs. The motion of the rail is eased by a set of bearings and it is affected by the rail length  and elasticity.

We look at the rail as a flexible body and use a finite elements approximation method to model the rail beams, resulting in a lumped-parameter model. This method assumes that beam deflection is small and in the linear regime. The lumped parameter approach approximates a flexible body as a set of rigid bodies coupled with springs and dampers. It can be implemented by a chain of alternating bodies and joints. The springs and dampers act on the bodies or the joints. The spring stiffness and damping coefficients are functions of the material properties and the geometry of the flexible elements.

\begin{figure*}[ht!]
\begin{center}
\includegraphics[width=0.8\textwidth]{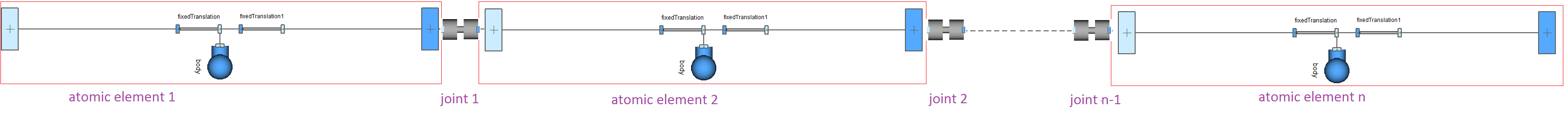}
\end{center}
\caption{Modelica rail model based on finite elements.}
\label{fig:03201542}
\end{figure*}

Each component of the rail-switch system was implemented in the Modelica language. Figure \ref{fig:03201542} shows the Modelica model of a beam used to model the rail.
The complexity of the rail switch and the rail Modelica models is shown in Table \ref{1:table}. We note that the majority of the model complexity is concentrated on the rail model. Hence producing a reduced representation of this model improves its usability, especially in real time applications.
\begin{table}[h!]
\centering
 \begin{tabular}{|c| c| c|c|c|c|}
 \hline
  & \textbf{No. of comp} & \textbf{Vars} & \textbf{Params} & \textbf{Diff. vars} & \textbf{Eqns}  \\
 \hline
 \textbf{Rail switch} & 595 & 8308 & 1574 & 336 & 5522 \\
 \hline
\textbf{Rail only} & 493 & 7244 & 1514 & 288 & 4768 \\
 \hline
 \end{tabular}
 \caption{Rail switch and rail model statistics}
  \label{1:table}
\end{table}
The input of the rail switch is a reference signal for the servo-motor controller for each of the two directions of motion. The time horizon  for each input reference signal is 7 sec. Using the high-fidelity model, it takes more than 9 sec to simulate the model over 14 sec during which the rail is pushed in both directions. Our objective is to replace the rail component with a simpler representation, to significantly reduce the simulation time, and enable real time diagnosis.

\section{Fault augmentation}
\label{sec_fault_augmentation}
\begin{figure}[ht!]
\begin{center}
\includegraphics[width=0.5\textwidth]{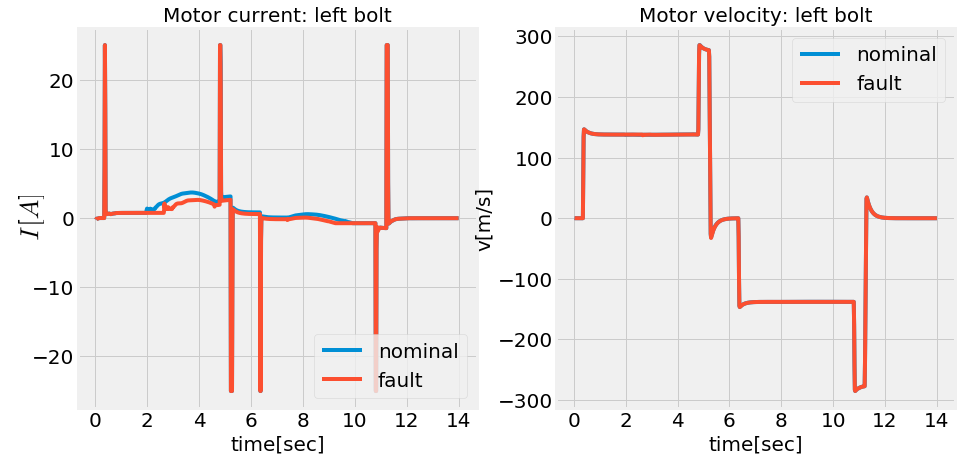}
\end{center}
\caption{Effects of a misaligned left adjuster bolt on the motor current and angular velocity.}
\label{fig:04141656}
\end{figure}

\begin{figure}[ht!]
\begin{center}
\includegraphics[width=0.5\textwidth]{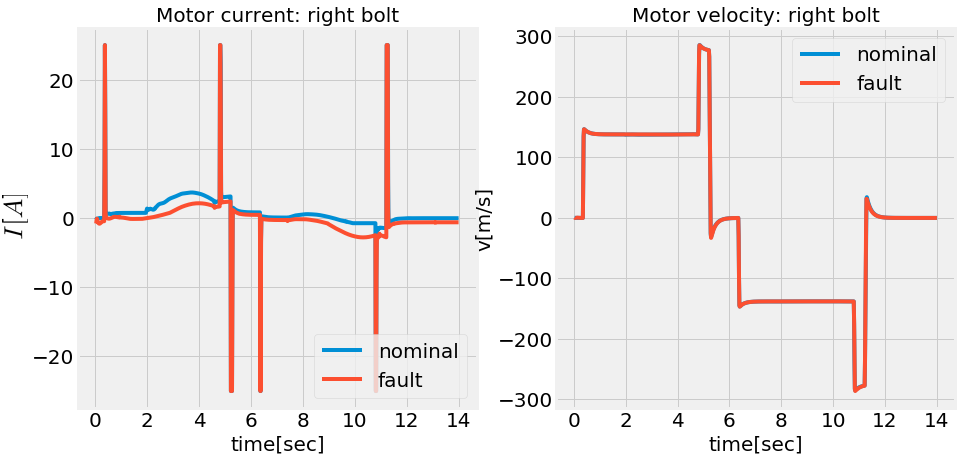}
\end{center}
\caption{Effects of a misaligned right adjuster bolt on the motor current and angular velocity.}
\label{fig:04141658}
\end{figure}

In this section we describe the modeling artifacts used recover  the behavior of the system for the four fault operating modes:  misaligned adjuster bolts (left and right), obstacle and missing bearings. These fault modes were reported to be of interest by a rail system operator we collaborated with. Obviously, there are many other fault modes that can originate from the point machine for example. Such faults are more readily detected due to the rich instrumentation present at the servo-motor.\\
\textbf{Misaligned adjuster bolts}: In this fault mode the bolts of the adjuster deviate from their nominal position. As a result, the instant at which the drive
rod meets the adjuster (and therefore the instant at which the switch rail starts moving) happens either earlier or later. For example, in a left-to-right motion, if the left bolt moves to the right, the contact happens earlier. The reason is that since the distance between the two bolts decreases, the left bolt reaches the adjuster faster. Hence, when the drive rod reaches its final position, there may be a gap between the right switch blade and the right stock rail. In contrast, if the left bolt moves to the left, the contact happens later. The model of the adjuster includes parameters that can set the positions of the bolts, and therefore the effects of this fault mode can be modeled without difficulty. Figures \ref{fig:04141656} and \ref{fig:04141658} show a comparison between the nominal behavior and the misaligned left and right bolts, respectively, in terms of the motor current and angular velocity.

\begin{figure}[ht!]
\begin{center}
\includegraphics[width=0.5\textwidth]{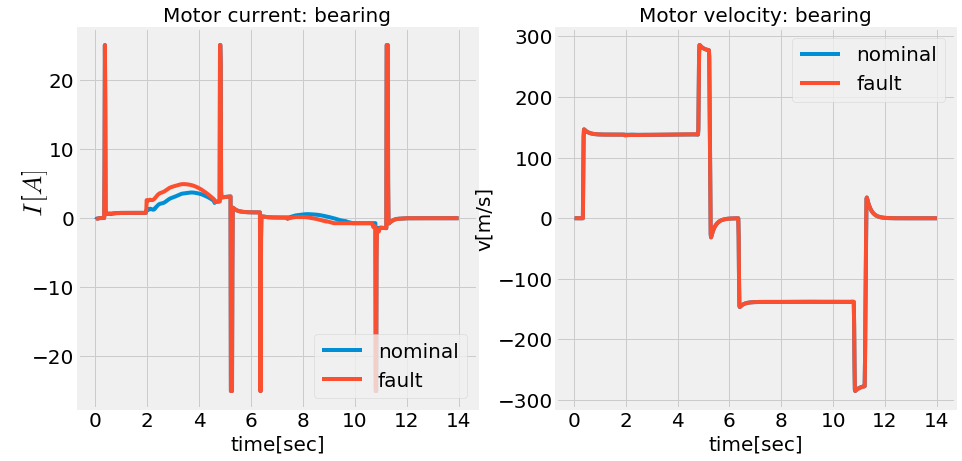}
\end{center}
\caption{Effects of  missing bearings on the motor current and angular velocity.}
\label{fig:04141700}
\end{figure}

\textbf{Missing bearings}:
To minimize friction, the rails are supported by a set of rolling bearings. When they become stuck or lost, the energy losses due to friction increase.  We included  a component connected to the rail to account for the additional friction. This component has a parameter that sets the value for the friction coefficient. By increasing the value of this parameter, the effect of the missing bearings fault can be simulated. A comparison between the motor current and angular velocity behavior under the nominal and missing bearing modes is shown in  Figure \ref{fig:04141700}.

\begin{figure}[ht!]
\begin{center}
\includegraphics[width=0.5\textwidth]{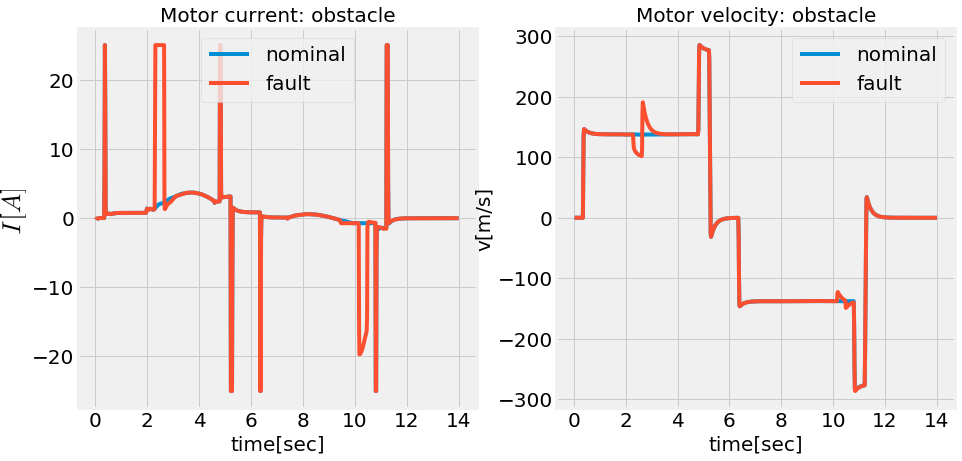}
\end{center}
\caption{Effects of  missing bearings on the motor current and angular velocity.}
\label{fig:04141704}
\end{figure}

\textbf{Obstacle}:
In this fault mode, an obstacle obstructs the motion of the switch blades. In case the obstacle is insurmountable, a gap between the switch blades and the stock rail appears. The effect on the motor current is a sudden increase in value, as the motor tries to overcome the obstacle. To model this fault we included a component that induces a localized, additional friction phenomenon for the switch blades. This component has two parameters: the severity of the fault and the position. For very high fault severity the switch blades cannot move beyond a certain position.  Figure \ref{fig:04141704} shows a comparison between the nominal behavior and the obstacle present behavior in terms of the motor current and angular velocity.

\section{Acausal modeling}
\label{sec_acausal_modeling}
Acausal models are physics based models typically constructed from first principles. Unlike the causal models used in signal processing and control, components of acausal models do not have inputs and outputs but ports (connectors) through which energy is exchanged with other components or with the environment. This is the modeling formalism used in the Modelica \cite{Fritzson15} language or in Simscape. Ports are characterized by  variables whose type determines how they are manipulated when two ports are connected. For example at a connection point, all flow variables sum up to zero (flow conservation), while all effort variables are equal. Examples of flow variables include current, force, torque while examples of effort variables include potential, velocity, angular velocity. Typically, the product between a flow and an effort variable has the physical interpretation of instantaneous power. The acausal modeling formalism is an instance of the more general port-Hamiltonian formalism \cite{h2}.
 The behavior of acausal components is determined by a set of constitutive equations of the form $f(x;w) = 0$, rather than by a causal map (with or without memory). The vector of variables $x$ can include port variables (flow, effort) and internal variables (states, algebraic variables), while $w$ is a vector of component parameters.

 Inspired by our previous work on hybrid modeling \cite{8431510,8814675}, we use acausal models to generate simplified representations of the rail component. To learn the parameters of the constitutive equations there are two main scenarios that can be considered. In the first scenario, we  assume that we can directly measure the component variables. This has the advantage that we can in theory perform the model learning in isolation, without considering the entire model. For this approach to work we need to carefully choose the model representation to avoid learning trivial models. The second scenario assumes we have only indirect information about the behavior of the component through measurements that do not include the rail component variables. In this case, the learning must include the entire rail-switch model and it is more computationally intense. Since we have access to the high fidelity model and hence we can directly measure every model variable, we  consider the first scenario. We use two type of representations for the (acausal) rail component: causal\footnote{We force a causal behavior for the acausal component.} and acausal.

 It the causal case we assume that some variables are inputs while other variables are outputs. This assumption is not ad-hoc. It comes from a causal analysis of the entire system model that produces causal relationships between the system variables. This causal analysis is typically performed before simulating a dynamical system represented as a DAE \cite{Casella11}. Once the input/output variable assignment is done, we select a representation for the constitutive equations (e.g., a neural network) and move to the parameter learning step.  Note that instead of assigning the component variables to an input/output category, we can try to learn the component parameters by assuming that all variables are inputs and the output is zero for all inputs. This approach can only work when considering the entire system model, case which introduces a regularization effect that prevents learning a trivial equation such as the constant zero map. Indeed, a zero map playing the role of a constitutive equation can make the system model unstable due to a singular Jacobian of the system DAE.

\begin{figure}[ht!]
\begin{center}
\includegraphics[width=0.3\textwidth]{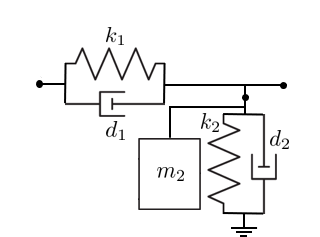}
\end{center}
\caption{Basic construct represented as compositions of generalized mass ($m$), spring ($k$), dampers ($d$).}
\label{fig:12281208}
\end{figure}

 In the acausal case the constitutive equations emulate physical laws. In what follows, we discuss different options for the constitutive equations that guarantee that the overall system model can be simulated. Since the behavior of the component can be fairly complex, we may need a large number of constitutive equations. To avoid arbitrary choices of constitutive equation maps, we use networks of generalized mass spring dampers (gMSD). In such a network, each  node is a composition of one generalized mass, spring and damper in a parallel connection, and each link is a composition of one spring and damper (see Figure \ref{fig:12281208}). To ensure that the component modeled as a network of gMSDs does not destabilize the overall system model, we impose conditions on the gMSDs that ensure that the model can be simulated. Such a condition is \emph{dissipativity}. A dissipative component cannot generate energy internally. A formal definition of a dissipative component is given in what follows.
\begin{definition}
Let $E(t) = E(t_0)-\int_{t_0}^tp(\tau) d\tau$ be the energy of a physical component, where $p(t)$ is its power. The component is \emph{dissipative} if $E(t)\leq E(t_0)$ for all $t\geq t_0$.
\end{definition}
We propose two type of maps for the gMSD that guarantees dissipativity. The first type if based on a polynomial representation as described in the following proposition.
\begin{proposition}
\label{pro:14030320}
Consider a component represented as a network of gMSD where the behavior of the masses, springs and dampers are given by
\begin{IEEEeqnarray*}{rCl}
\label{eq1:12111038}
  F_m &=& \sum_{i=0}^n m_i \textmd{sign}(\dot{x})\ddot{x}^i,F_c = \sum_{i=0}^n c_i \textmd{sign}(\dot{x}){x}^i\\
  \label{eq1:12111039}
  F_d &=& \sum_{i=0}^n d_i \textmd{sign}(\dot{x})\dot{x}^i,
\end{IEEEeqnarray*}
respectively, where the scalars $m_i$, $d_i$ and $c_i$ are non-negative scalars, and $n$ is the polynomial order. Then the component is dissipative.
\end{proposition}
An alternative definition for the gMSD is given in the following proposition.
\begin{proposition}
\label{pro:14070320}
Consider a component represented as a network of gMSD where the behavior of the masses, springs and dampers are given by:
\begin{equation*}
F_m =  m(x,\dot{x},\ddot{x})\ddot{x}, F_c = k(x)x, F_d = d(x,\dot{x})\dot{x},
\end{equation*}
respectively, where  $m(\cdot,\cdot,\cdot)$, $k(\cdot)$ and $d(\cdot,\cdot)$ are bounded positive maps. Then the component is dissipative.
\end{proposition}

Note that we have a lot of freedom with respect of modeling the functions  $m(\cdot,\cdot,\cdot)$, $k(\cdot)$ and $d(\cdot,\cdot)$. We can model them for example as neural networks, where we make sure that the last layer imposes a non-negative output through a ``ReLu'' layer or by taking the square of the output of the last linear layer.
Since the constitutive equations may contain differential equations, we will need to use learning platforms with ODE (e.g., Pytorch \cite{paszke2017automatic}, TensorFlow \cite{tensorflow2015-whitepaper}) or DAE (e.g., DAETools \cite{10.7717/peerj-cs.54}) solving capabilities, if the state derivatives are not measured.

\section{Hybrid rail switch model}
\label{sec_hybrid_rail_switch_model}

\begin{figure}[ht!]
\begin{center}
\includegraphics[width=0.48\textwidth]{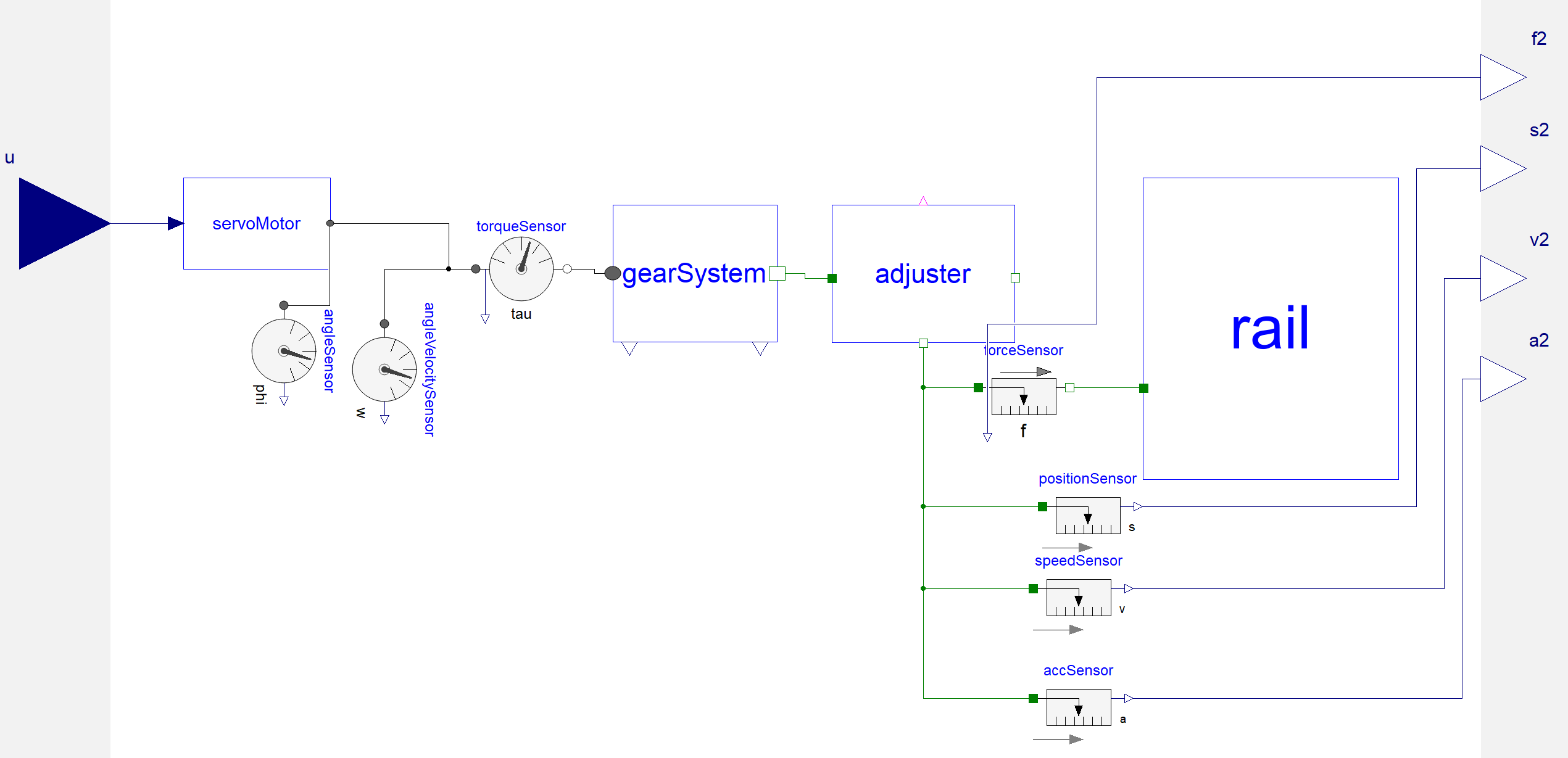}
\end{center}
\caption{High fidelity Modelica model for  rail switch.}
\label{fig:03201554}
\end{figure}

\begin{figure}[ht!]
\begin{center}
\includegraphics[width=0.48\textwidth]{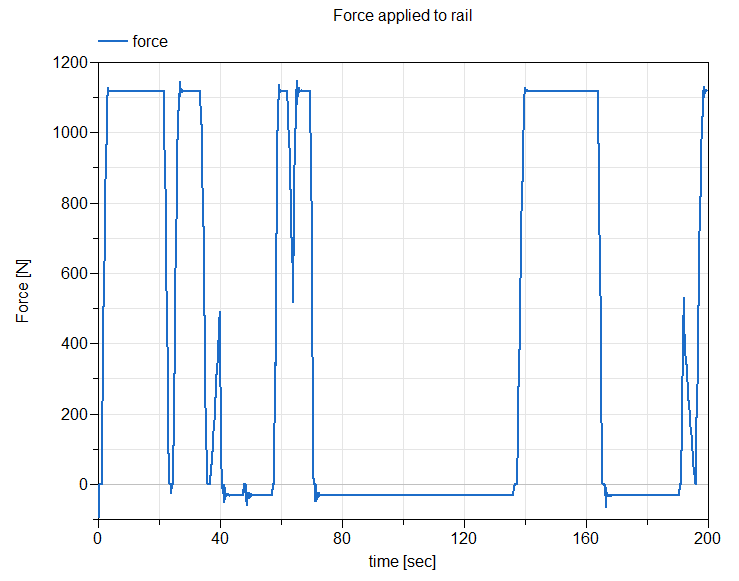}
\end{center}
\caption{Force applied to the high fidelity rail model.}
\label{fig:03201834}
\end{figure}

\begin{figure}[ht!]
\begin{center}
\includegraphics[width=0.48\textwidth]{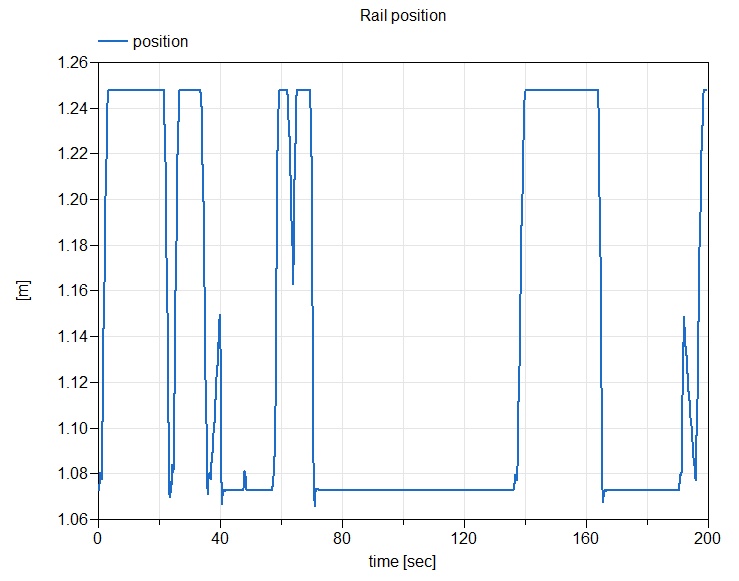}
\end{center}
\caption{Position of the high fidelity rail model.}
\label{fig:03201835}
\end{figure}

\begin{figure}[ht!]
\begin{center}
\includegraphics[width=0.48\textwidth]{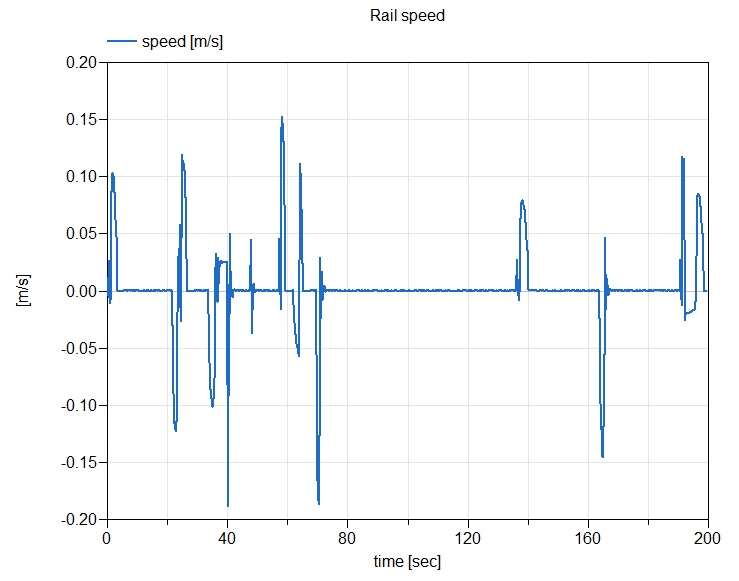}
\end{center}
\caption{Speed of the high fidelity rail model.}
\label{fig:03201836}
\end{figure}

In this section we introduce several approaches for simplifying the rail switch component model. In addition to model simplification, we will also focus on preserving the physical interpretation of the reduced model, through appropriate choices of constitutive equations. We assume that  we have access to the variables at the connection point between the adjuster and the rails. In particular, we assume we can directly measure the force $F$, position $x$, velocity $v$ and acceleration $a$. We use the two modeling approaches:
\begin{itemize}
\item \emph{Causal approach}: we determine a causal relation between the force, position, velocity an acceleration and use a causal map such as a neural network to model the relation between them. The resulting component model is still acausal though, with an imposed variable dependence.
    \item \emph{Acausal approach}: we model the rail component as a combination of generalized mass, spring, dampers as defined in Propositions \ref{pro:14030320} and \ref{pro:14070320}. We will show that one mass-spring-damper component is sufficient.
\end{itemize}
The training data is generated by simulating the high fidelity rail switch model. The block diagram of the rail switch Modelica model is shown in Figure \ref{fig:03201554}. It has as input current references for the serve-motor, correlated with a desired velocity profile. Typically, pre-determined current trajectories are fed to the servo-motor to generate the rail motion. In our case, we will use random inputs to push the rail. We will record the force, position, velocity and acceleration trajectories and use them as training data. Each time series corresponds to a time interval of 100 sec, sampled at 0.1 sec. When appropriate, we use one time series for training or several of them.
Examples of rail force, position and speed trajectories used for training are shown in Figures \ref{fig:03201834}, \ref{fig:03201835} and \ref{fig:03201836}, respectively.
\subsection{Causal modeling}
\begin{figure}[ht!]
\begin{center}
\includegraphics[width=0.48\textwidth]{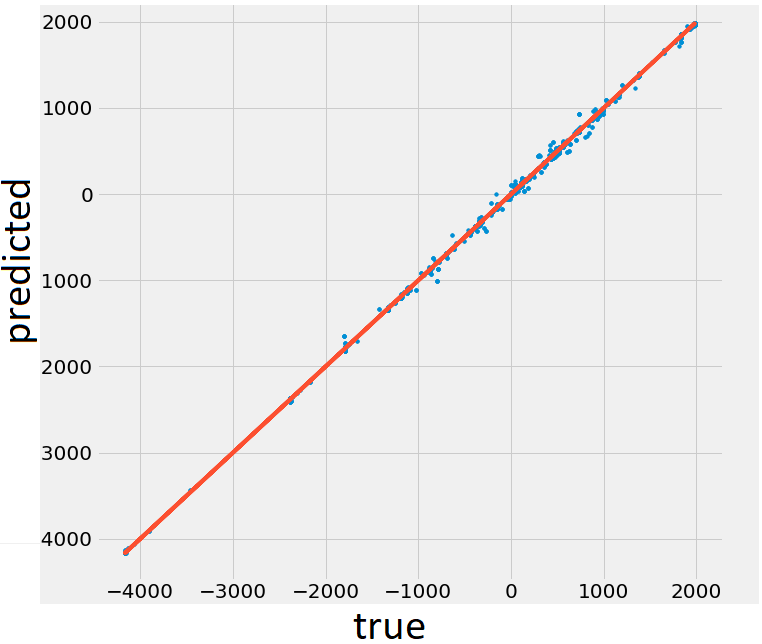}
\end{center}
\caption{Validation of the learned model.}
\label{fig:03211049}
\end{figure}

In this approach, we assign causality relations to the variables at the connection point between the adjuster and the rails. Since the servo-motor tracks a pre-specified speed pattern, our intuition should tell us that the position and velocity of the rails are set by the motor. This intuition is confirmed by a causal analysis performed by looking at the block lower triangular (BLT) transformation \cite{Casella11} that depicts the causal relations between the system variables\footnote{The BLT transformation is too large to be included in a plot.}. Hence, we model the rail behavior by using a causal map $F = g(u;w)$, where $g:\mathds{R}^3\rightarrow \mathds{R}$ is a map described by a neural network (NN) with one hidden layer:
$$g(u) = W^{[1]}\left(\tanh\left(W^{[0]}u+b^{[0]}\right)\right)+b^{[1]},$$
where, the input $u = [x,\dot{x},\ddot{x}]$ is a vector containing the position, speed and acceleration, the output $F$ is the force,  and $w=\{W^{[0]},b^{[0]},W^{[1]},b^{[1]}\}$ is the set of parameters of the map $g$.
We have employed a two step training process. In the first step we train the parameters of the map in isolation, considering the map $g$ only. We  used $15$ time series containing trajectories of the force, position, speed and acceleration. We used the Keras \cite{chollet2015keras} deep-learning training platform, proceeded by  splitting the data into training ($70\%$) and test ($30\%$) data sets. We chose the hidden layer dimension to be 50, and trained the NN parameters using a decaying learning rate. The validation results are shown in Figure \ref{fig:03211049}, where we depict the true vs. predicted output samples using as input the test data set. The MSE for the validation data is $MSE_{test} = 415.46$. Although it may appear a large value, it must be interpreted relative to the values of the force used in training and validation, since the training data was not normalized to maintain physical interpretation.
We used the weights of the Keras model to implement a Modelica component with one port and the constitutive equation given by $F = W^{[1]}\left(\tanh\left(W^{[0]}u+b^{[0]}\right)\right)+b^{[1]}$, where $u = [x,\dot{x},\ddot{x}]$.
Next, we executed a fine tuning of the component parameters by performing a parameter learning step using the entire rail switch model. This way, the rest of the model equations are considered, adding an additional regularization effect. We chose a gradient-free optimization algorithm, namely the Powell algorithm, to avoid using gradient approximations. The Modelica rail switch model was converted into a functional mockup unit (FMU) \cite{Blochwitz11thefunctional}, and integrated with the Powell algorithm in Python. Although  gradient free algorithms are typically slow for a large number of variables, we did not have to run the algorithm for a large number of iterations since we used the Keras solution as initial parameter values. The result of this additional step was a $20\%$ improvement of the loss function applied to the test data.  The complexity of the Modelica component is shown in Table \ref{2:table}.

{\small \begin{table}[h!]
\centering
 \begin{tabular}{|p{1.7cm}| p{1cm}| p{1cm}| p{1.2cm}|p{1.2cm}|}
 \hline
  & \textbf{No. of components} & \textbf{Variables} & \textbf{Parameters}  & \textbf{Equations}  \\
 \hline
 \textbf{Causal rail representation} & 1 & 15 & 255  & 8 \\
 \hline
 \end{tabular}
 \caption{Complexity of the Modelica rail model using a causal map.}
  \label{2:table}
\end{table}}

\subsection{Acausal modeling}
\begin{figure}[ht!]
\begin{center}
\includegraphics[width=0.48\textwidth]{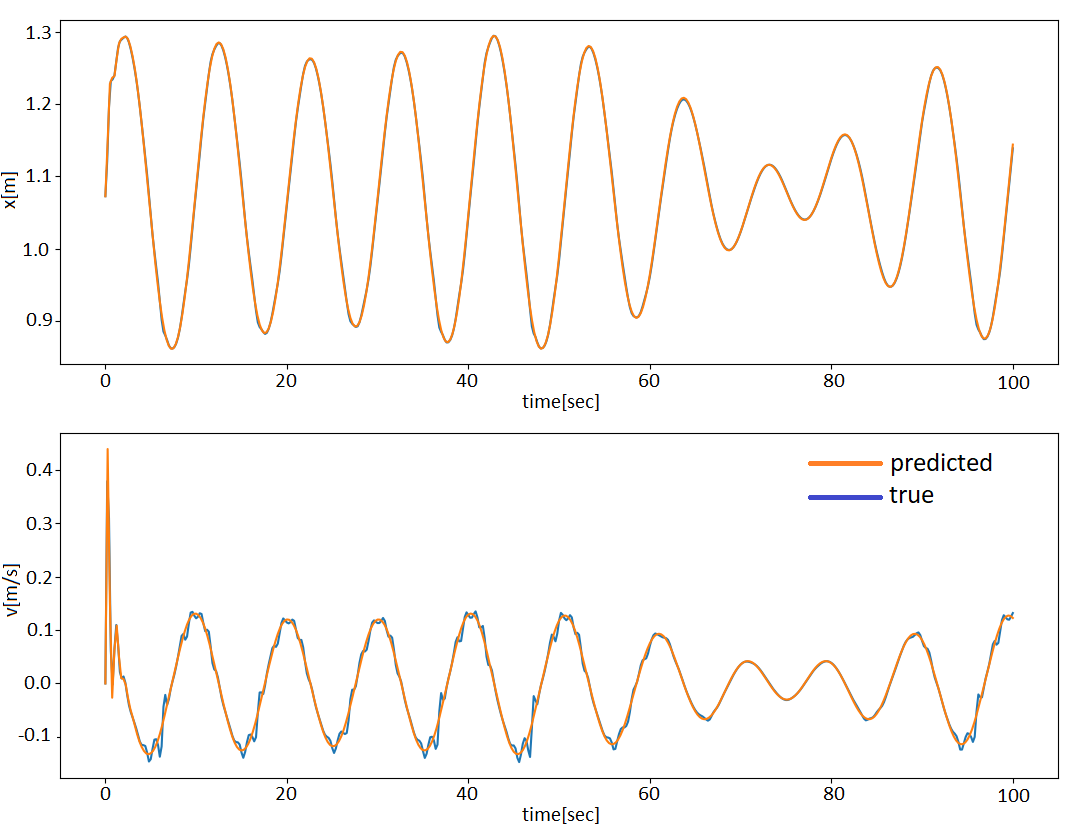}
\end{center}
\caption{True (blue) vs predicted (orange) position and speed time series for the acausal rail model in polynomial form.}
\label{fig:0321243}
\end{figure}

We showed in the previous section how we can use causal maps inside acausal components. The advantage of the causal representation is that we can use main stream deep learning platforms to learn the parameters of the causal map. There is a significant disadvantage though: it is not clear if the obtained component is reusable. By reusability we understand the ability to use the component in different configurations and still behaving as expected. From a numerical perspective view, this means that we should be able to compute the acceleration when the force becomes the input (position and speed are state variables and considered known from the previous simulation step). The acausal modeling approach guarantees this. Using the observation that the rail opposes motion, we modeled the rail as a combination of a generalized mass-spring damper in a parallel connection. We use two types of gMSD models: polynomial and NN. We considered a linear mass model: $F_m = m\ddot{x}$ for both cases. In the polynomial case, we considered the following models for the spring and damper, respectively: $F_c = c_0(x-x_{fix})+c_1(x-x_{fix})^3+c_2(x-x_{fix})^5$ and $F_d = d_0\dot{x}+d_1\dot{x}^3+d_2\dot{x}^{5}$. The set of parameters we have to learn is $w=\{m,c_0,c_1,c_2,d_0,d_1,d_2,x_{fix}\}$.  Unlike to previous section, we considered as input the force, and as outputs the position and velocity. The model parameters are the solution of the following constrained optimization problem:
\begin{IEEEeqnarray*}{rCl}
\min_{w\geq 0} & & \frac{1}{2N}\sum_{i=1}^N\|x(t_i)-\hat{x}(t_i)\|^2+\|\dot{x}(t_i)-\hat{\dot{x}}(t_i)\|^2\\
\textmd{subject to: }& &\\
& & m\hat{\ddot{x}}(t_i) + F_c(t_i)+F_d(t_i) = F(t_i),\\
& &F_c(t_i) = c_0(\hat{x}(t_i)-x_{fix})+c_1(\hat{x}(t_i)-x_{fix})^3\\
& &+c_2(\hat{x}(t_i)-x_{fix})^5,\\
& &F_d(t_i) = d_0\hat{\dot{x}}(t_i)+d_1\hat{\dot{x}}(t_i))^3+d_2\hat{\dot{x}}(t_i)^5,\\
& &w=\{m,c_0,c_1,c_2,d_0,d_1,d_2,x_{fix}\}.
\end{IEEEeqnarray*}
where $t_i$ are time samples of the time series. The optimization problem used one time series only and used a nonlinear least square algorithm. We relied on the  DAETools \cite{10.7717/peerj-cs.54} Python package to implement the optimization algorithm since it provides access to the gradients of the cost function, hence gradient  approximations are not needed. The resulting optimal parameters are as follows: $c_0^* = 6.5\times 10^{3}$, $c_1^*=0.45$, $c_2^* = 4.15\times 10^{4}$, $d_0^* = 5.96\times 10^2$, $d_1^* = 0$, $d_2^*=0$, $m^*=1.5\times 10^{2}$, $s_0^*= 1.077$. The ``true'' vs predicted time series are show in Figure \ref{fig:0321243}.
The learning result shows that the polynomial representation for the acausal rail model captures the dominant behavior of both the position and speed.
We repeated the learning process when the acausal rail model is represented using NN representations. In particular we chose as models for the spring and damper $F_c=c(x,\dot{x})^2(x-x_{fix})$ and $F_d = d(x,\dot{x})^2\dot{x}$, respectively,  where $c(x,\dot{x})$ and $d(x,\dot{x})$ are modeled as neural networks with one hidden layer of size 15 and {\tt tanh} as activation function. Using  the DAETool, we solved the following optimization problem:

\begin{IEEEeqnarray*}{rCl}
\min_{w} & &\frac{1}{2N}\sum_{i=1}^N\|x(t_i)-\hat{x}(t_i)\|^2+\|\dot{x}(t_i)-\hat{\dot{x}}(t_i)\|^2\\
\textmd{subject to: } &  & m\hat{\ddot{x}}(t_i) + F_c(t_i)+F_d(t_i) = F(t_i),\\
& & F_c(t_i) = c(\hat{x}(t_i),\hat{\dot{x}}(t_i))^2(\hat{x}(t_i)-x_{fix}),\\
& & F_d(t_i) = d(\hat{x}(t_i),\hat{\dot{x}}(t_i))^2\hat{\dot{x}}(t_i),\\
& & c(\hat{x}(t_i),\hat{\dot{x}}(t_i)) =\\
& & W_c^{[1]}\tanh(\left(W_c^{[0]}[x(t_i),\dot{x}(t_i)]^{T}+b_c^{[0]}\right))+b_c^{[1]},\\
& & d(\hat{x}(t_i),\hat{\dot{x}}(t_i)) =\\
& & W_d^{[1]}\tanh(\left(W_d^{[0]}[x(t_i),\dot{x}(t_i)]^{T}+b_d^{[0]}\right))+b_d^{[1]},\\
& & w=\{m,W_c^{[1]},b_c^{[1]},W_c^{[0]},b_c^{[0]},W_d^{[1]},b_d^{[1]},\\
& & W_d^{[0]},b_d^{[0]},x_{fix}\}
\end{IEEEeqnarray*}

 \begin{figure}[ht!]
\begin{center}
\includegraphics[width=0.48\textwidth]{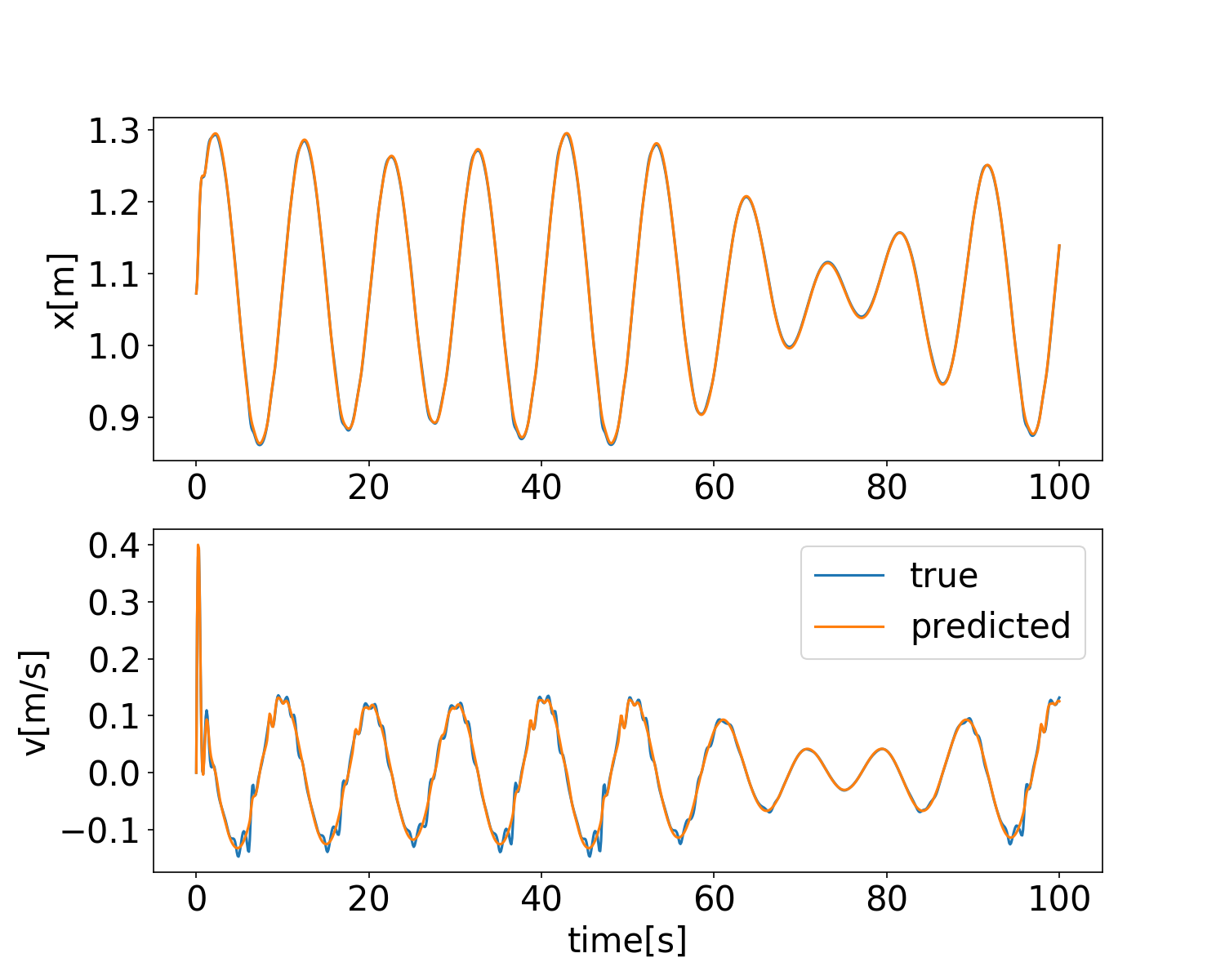}
\end{center}
\caption{True (blue) vs predicted (orange) position and speed time series for the acausal rail model in NN form.}
\label{fig:0321303}
\end{figure}
As seen in Figure \ref{fig:0321303}, with the neural network representation we are able to recover a more detailed behavior for the speed. The complexity of the acausal rail models in the two representations are shown in Table \ref{3:table}.

\begin{figure}[ht!]
\begin{center}
\includegraphics[width=0.48\textwidth]{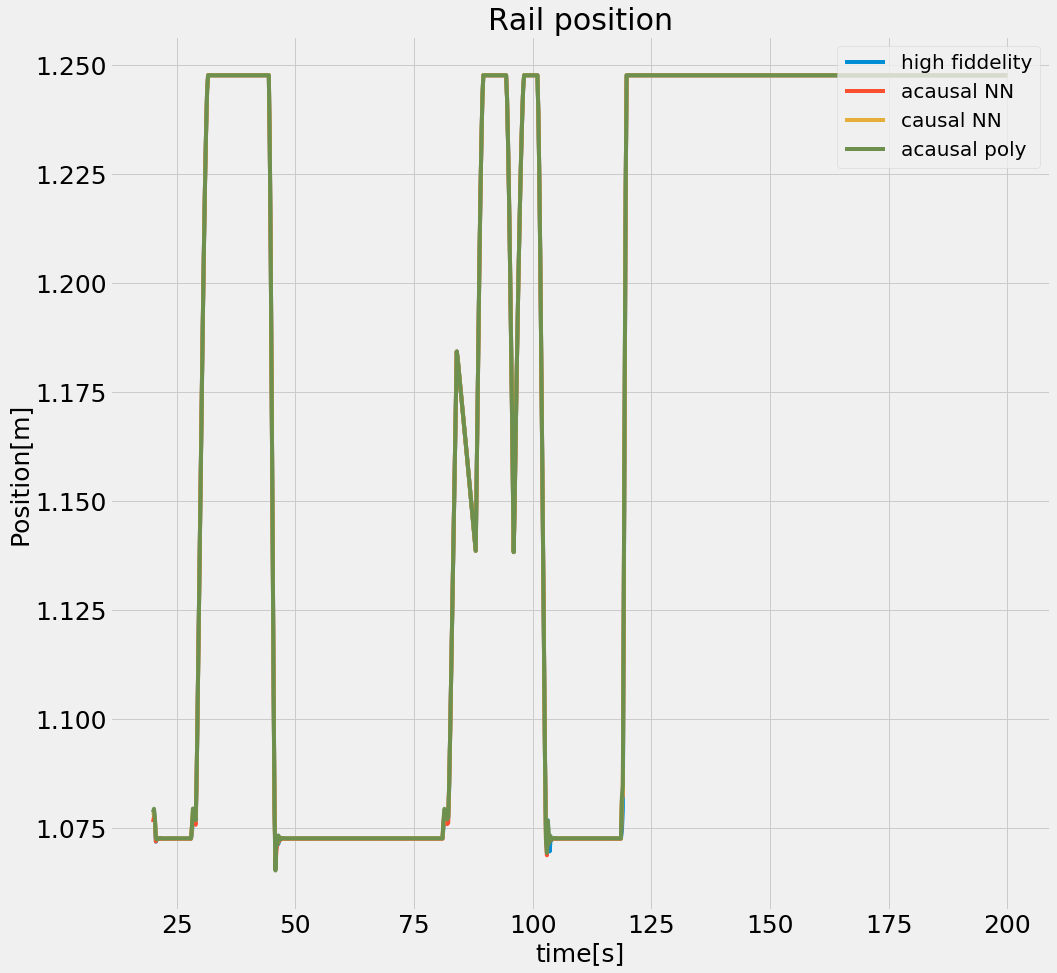}
\end{center}
\caption{Rail position for the high and low fidelity models.}
\label{fig:03221050_}
\end{figure}

\begin{figure}[ht!]
\begin{center}
\includegraphics[width=0.48\textwidth]{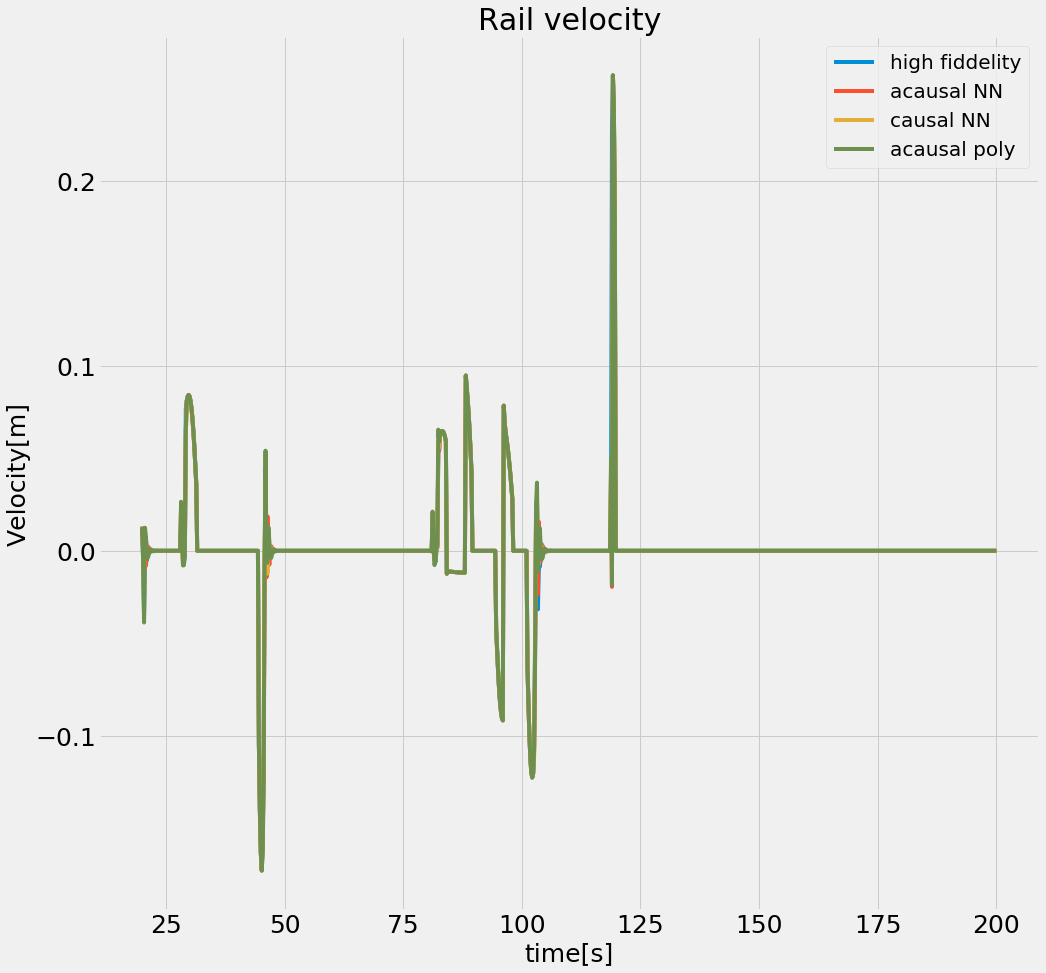}
\end{center}
\caption{Rail velocity for the high and low fidelity models.}
\label{fig:03221051_}
\end{figure}

\begin{figure}[ht!]
\begin{center}
\includegraphics[width=0.48\textwidth]{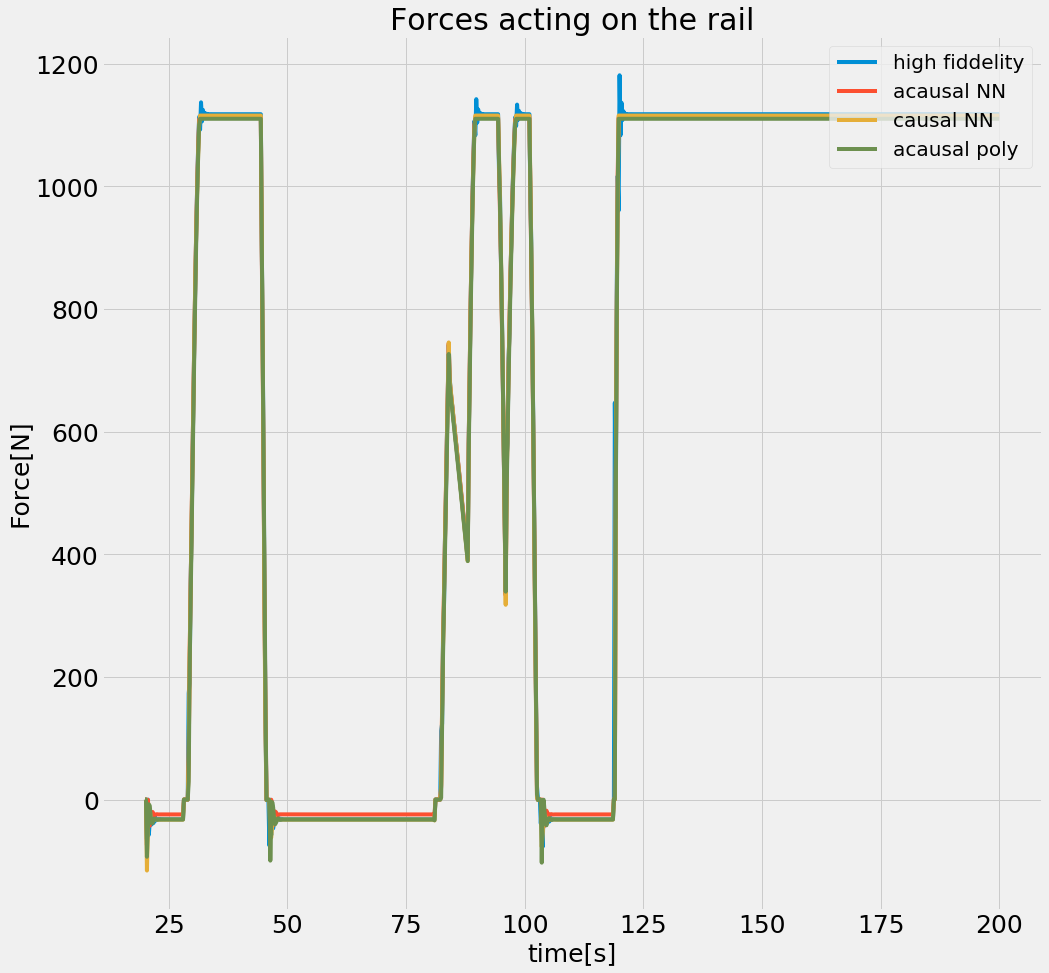}
\end{center}
\caption{Rail force for the high and low fidelity models.}
\label{fig:03221052_}
\end{figure}

\begin{table}[h!]
\centering
 \begin{tabular}{|p{1.7cm}| p{1cm}| p{1cm}|p{1.2cm}|p{1.2cm}|}
 \hline
  & \textbf{ No. of components } & \textbf{ Variables } & \textbf{ Parameters }  & \textbf{ Equations }  \\
 \hline
 \textbf{ Acausal poly } & 1 & 17 & 9  & 7 \\
 \hline
  \textbf{ Acausal NN } & 1 & 25 & 128 & 11 \\
 \hline
 \end{tabular}
 \caption{Complexity of the Modelica rail model using acausal representations}
  \label{3:table}
\end{table}
We validated the learned models by integrating them within the overall rail switch model. We generated 25 time series with random inputs for the servo-motor used for the four rail switch models: the high fidelity one, and three low fidelity corresponding to the causal NN, acausal polynomial and acausal NN representations, respectively. An example of such time series is shown in Figures \ref{fig:03221050_}, \ref{fig:03221051_} and \ref{fig:03221052_}.

\begin{figure}[ht!]
\begin{center}
\includegraphics[width=0.48\textwidth]{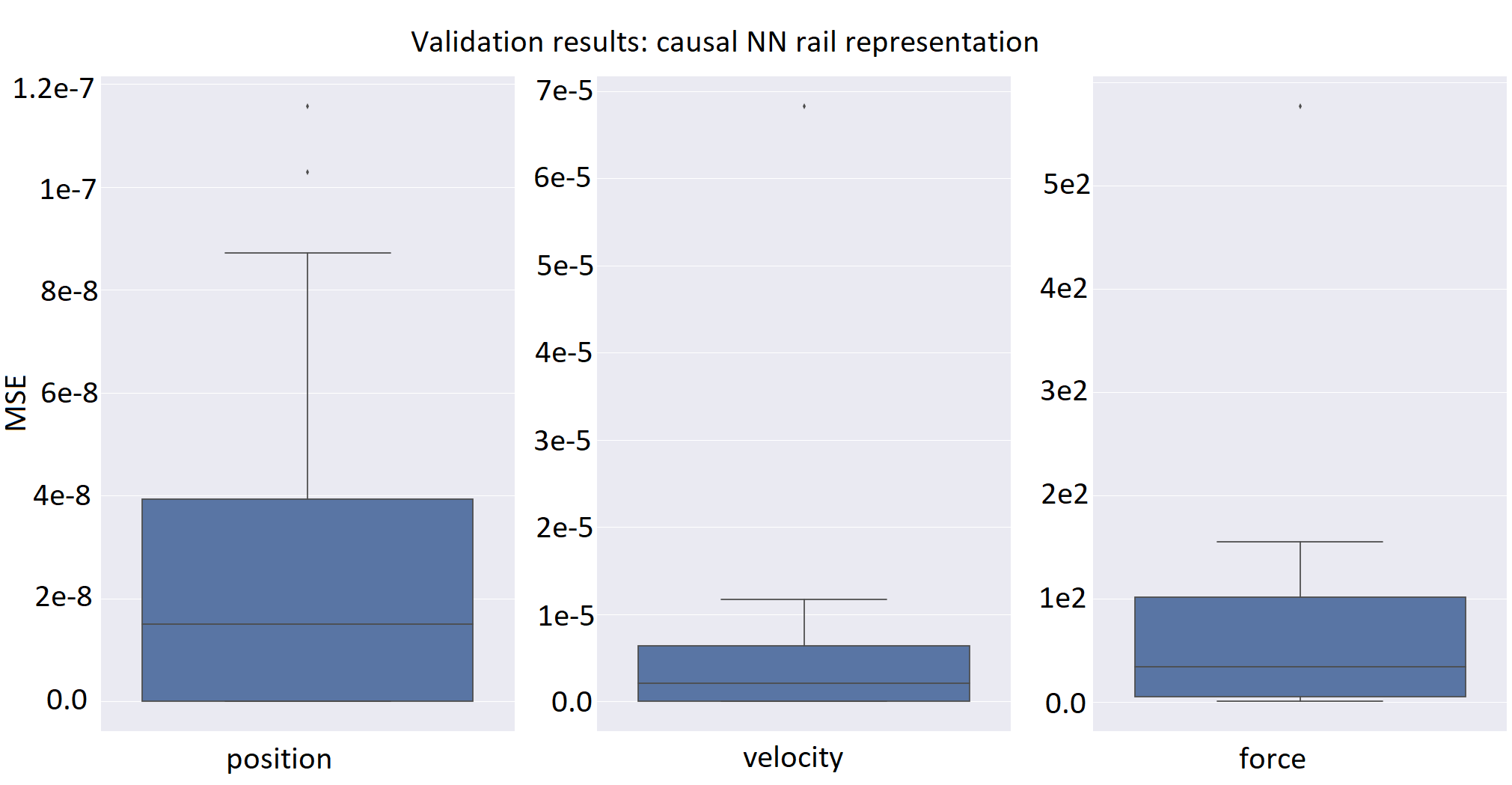}
\end{center}
\caption{Validation statistical results: causal rail representation.}
\label{fig:0321325}
\end{figure}

\begin{figure}[ht!]
\begin{center}
\includegraphics[width=0.48\textwidth]{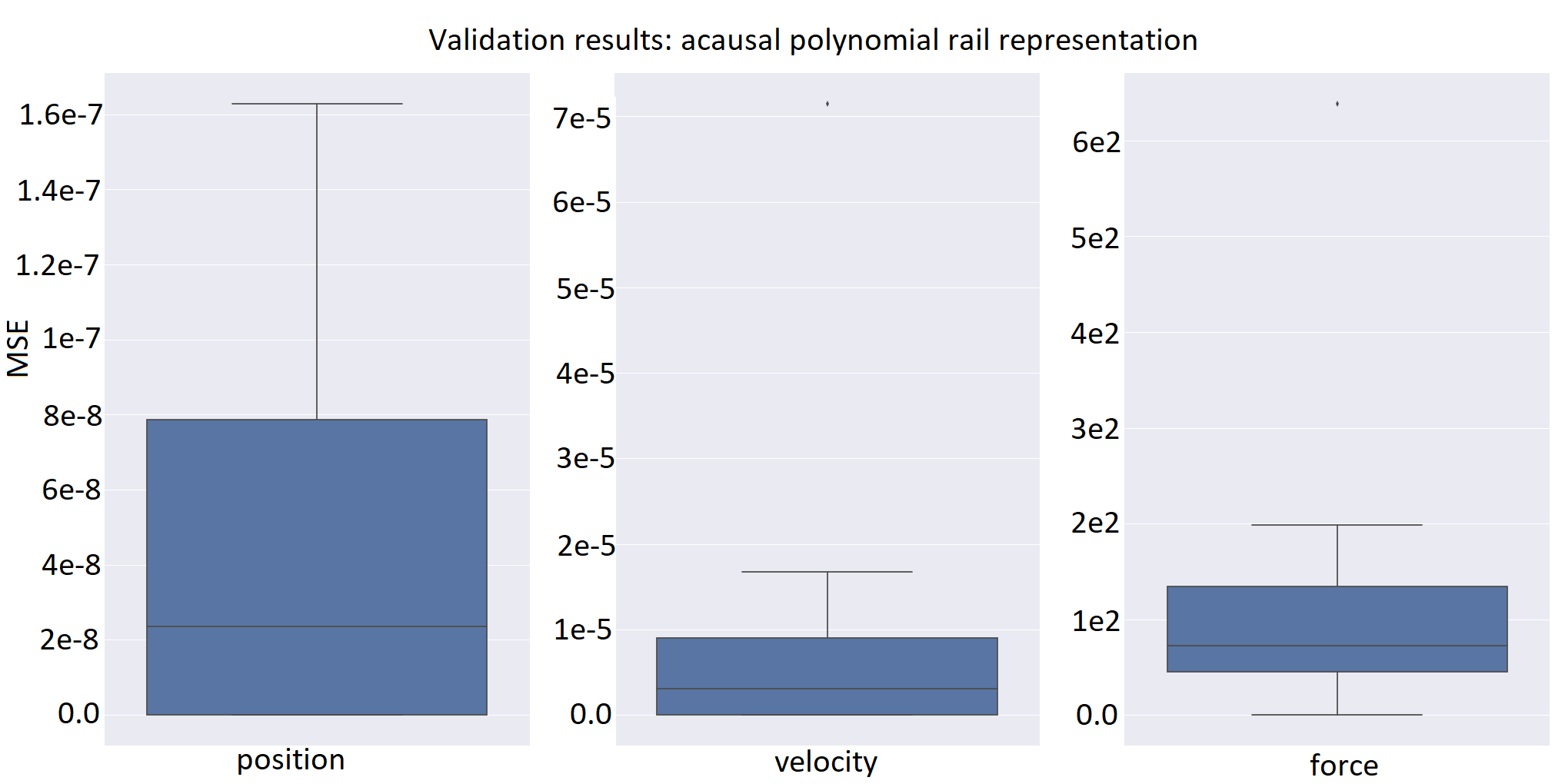}
\end{center}
\caption{Validation statistical results: acausal polynomial rail representation.}
\label{fig:0321326}
\end{figure}

\begin{figure}[ht!]
\begin{center}
\includegraphics[width=0.48\textwidth]{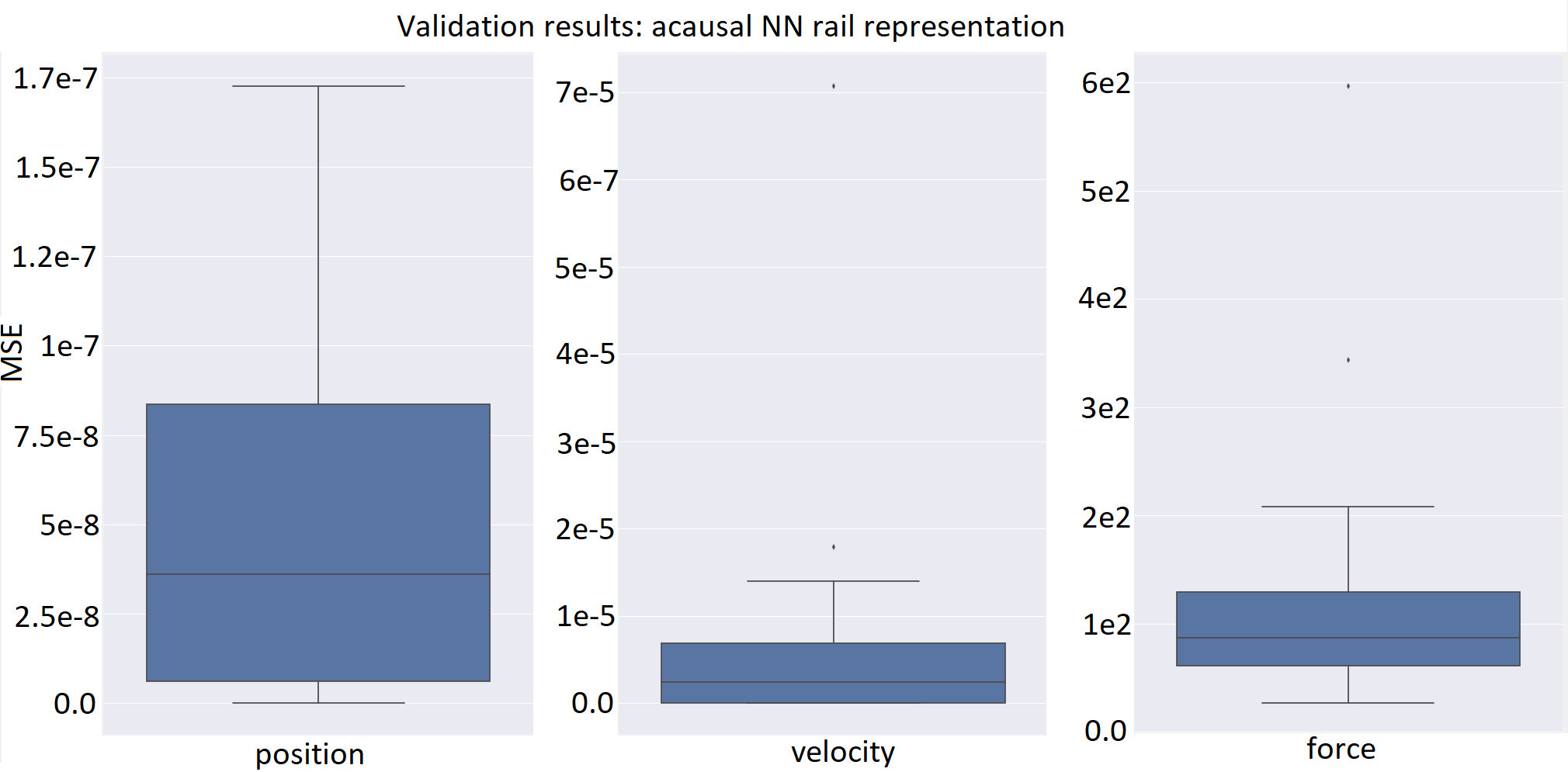}
\end{center}
\caption{Validation statistical results: acausal NN rail representation.}
\label{fig:0321327}
\end{figure}

We used the 25 time series to compute MSE statistics for the position, velocity and force. The results are shown in Figures \ref{fig:0321325}, \ref{fig:0321326} and \ref{fig:0321327}, shown as box plots. A first observation is that the MSEs corresponding to the force have large values as compared to the position and velocity. This should not be a surprise since the absolute values of the force are in the thousands. The position and velocity MSEs are similar for all three cases. In the case of the force, the acausal representations have roughly the same statistics, while in the causal case, the MSE has both the variances and mean comparable, but slightly smaller. This again should not be a surprise since the rail causal model is tailored for our scenario. In other works the model may be overfitted. In a different usage scenario, the casual representation may not even simulate. Hence we have a trade-off between accuracy and generalizability.

 \begin{figure}[ht!]
\begin{center}
\includegraphics[width=0.48\textwidth]{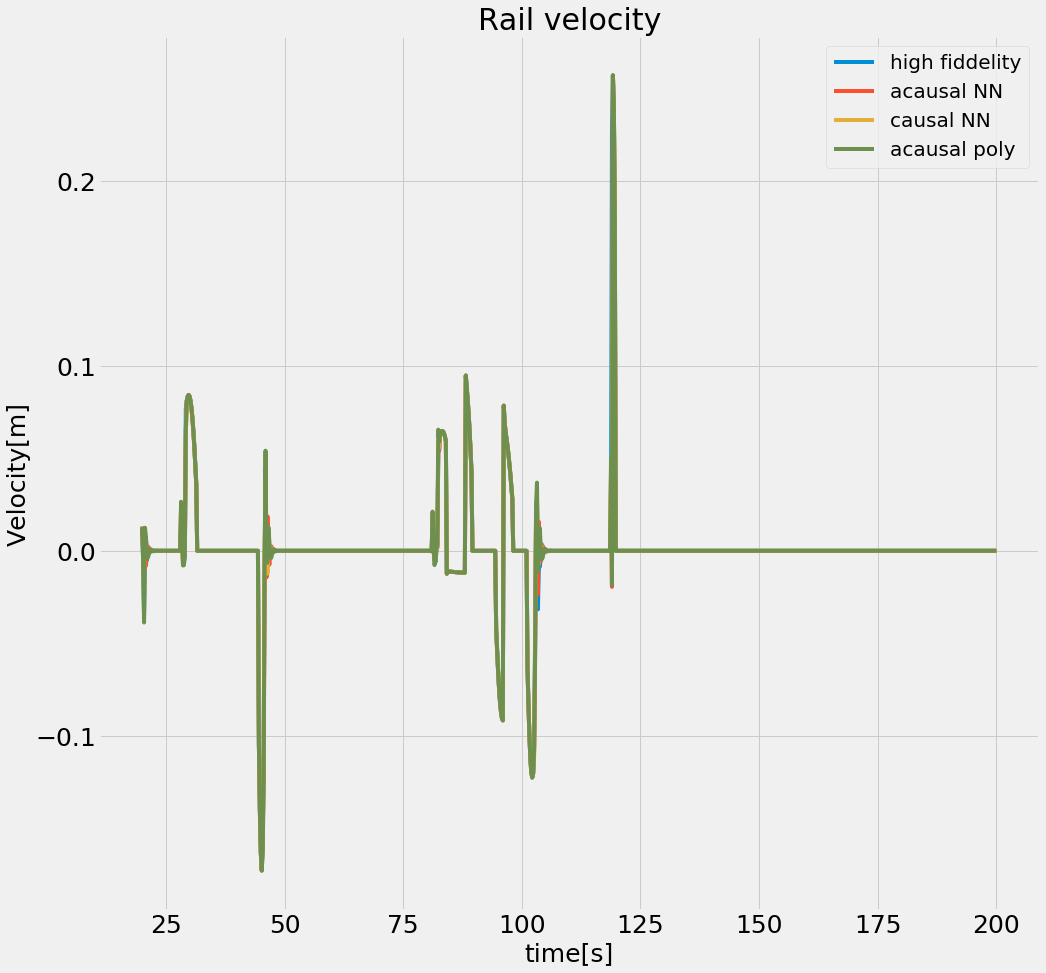}
\end{center}
\caption{Rail velocity for the high and low fidelity models.}
\label{fig:03221051}
\end{figure}

\section{Fault diagnosis}
\label{sec_fault_diagnosis}

\subsection{Optimization-based parameter estimation}
We estimated the fault parameters for each of the four fault modes using an optimization-based parameter estimation algorithm. The loss function was defined as the mean square error (MSE) between the simulated variables and the ``observed'' variables (motor current, motor angle and angular velocity). The observed variables were generated using the high-fidelity models and contain simulations over a time horizon of 14 sec including both switch motions: left to right and right to left. The variables are sampled at 0.1 sec. The optimization algorithm requires  loss function evaluations that in turn requires model simulations. The model simulations were done using Functional Mockup Units (FMU) \cite{Blochwitz11thefunctional} representations of the Modelica models. We tested the optimization algorithm for the three versions of the reduced complexity models: causal NN, acausal NN and acausal polynomial representations of the rail. We  tested the parameter estimation using several optimization algorithm including gradient-based and gradient-free algorithm. The best results were produced by the the differential evolution algorithm and they are presented in what follows. Since such an algorithm requires many loss function evaluations it is imperative for the model simulations to be fast. In average, the acausal polynomial, acausal NN and causal NN representations take 0.3 sec, 0.5 sec and 0.9 sec over the 14 sec time horizon. For the same time interval, the high fidelity model takes 7 sec. The FMUs were used in Python scripts implementing the parameter estimation algorithms. The model simulations were performed on PC with Intel 12 cores Xeon 3.5 GHz CPU with 64 GB of RAM.  We recall that the starting position of the rail is 1 m, value dictated by the initial conditions of the motor and the positions of the different reference points in the rail model.\\
\textbf{Left bolt fault}: the left bolt fault mode is active with a deviation from its nominal value of 50 mm. Tables \ref{1:table}-\ref{3:table} present the results of the parameter estimation algorithms corresponding to the three representations, when tracking separately the fault parameters. Using as metric the MSE it is clear that we are correctly identifying the left bolt fault as the current fault mode. In addition, the fault parameter values are within 3\%  of the value used to generate the faulty behavior.

\begin{table}[h!]
\centering
 \begin{tabular}{|c| c|c|}
 \hline
  \textbf{Tracked fault parameters}& \textbf{Paramater value} &  \textbf{MSE} \\
 \hline
 \textit{Left bolt}[mm] & \cellcolor{blue!25} 49.53  & \cellcolor{blue!25} 0.006 \\
 \hline
\textit{Right bolt}[mm] & 6.03  & 0.324 \\
 \hline
\textit{Missing bearing}[Ns/m] & 39.73 & 0.326 \\
 \hline
\textit{Obstacle}\{[Ns/m],[m]\} & \{1.5$\times 10^5$, 1.49\}  & 0.334  \\
 \hline
 \end{tabular}
 \caption{Left bolt fault mode (acausal polynomial representation).}
  \label{1:table}
\end{table}

\begin{table}[h!]
\centering
 \begin{tabular}{|c| c|c|}
 \hline
  \textbf{Tracked fault parameters}& \textbf{Paramater value} &  \textbf{MSE} \\
 \hline
 \textit{Left bolt}[mm] & \cellcolor{blue!25}48.69  & \cellcolor{blue!25}0.011 \\
 \hline
\textit{Right bolt}[mm] & 11.26  & 0.352 \\
 \hline
\textit{Missing bearing}[Ns/m] & 28.45 & 0.353 \\
 \hline
\textit{Obstacle}\{[Ns/m],[m]\}  & \{1.67$\times 10^5$, 10.04\}  & 0.386  \\
 \hline
 \end{tabular}
 \caption{Left bolt fault mode (acausal NN representation).}
  \label{2:table}
\end{table}

\begin{table}[h!]
\centering
 \begin{tabular}{|c| c|c|}
 \hline
  \textbf{Tracked fault parameters}& \textbf{Paramater value} &  \textbf{MSE} \\
 \hline
 \textit{Left bolt}[mm] & \cellcolor{blue!25}50.42  & \cellcolor{blue!25}0.005 \\
 \hline
\textit{Right bolt}[mm] & 8.91  & 0.344 \\
 \hline
\textit{Missing bearing}[Ns/m] & 33.67 & 0.304 \\
 \hline
\textit{Obstacle}\{[Ns/m],[m]\} & \{6.31$\times 10^4$, 0.0772\}  & 0.341 \\
 \hline
 \end{tabular}
 \caption{Left bolt fault mode (causal NN representation).}
  \label{3:table}
\end{table}

We estimated simultaneously all the fault parameters as well. The results for the three representations of the rail model are shown  in Table \ref{4:table}. We obtained reasonable small MSE values, but it is more challenging to distinguish between the faults modes. Recalling that the obstacle was introduced at 1.1 meters we can exclude the obstacle fault mode (the fault intensity is irrelevant outside the obstacle position). The parameter corresponding to the missing bearing fault mode has a value in the hundreds for two of the rail representations. Although they may appear not to have a significant impact on the behavior of the rail switch, without some prior information about what is a significant value it is difficult to draw a conclusion about this fault mode. The good news is that the left bolt fault parameter was reasonably well estimated. Although not zero, the right bolt fault parameter values are small enough to eliminate this fault mode as a possible source of faulty behavior.

\begin{table}[h!]
\centering
 \begin{tabular}{|p{2.24cm}| p{1.5cm}|p{1.5cm}|p{1.5cm}|}
 \hline
  \textbf{Tracked fault parameters}& \textbf{Acausal poly} & \textbf{Acausal NN} & \textbf{Causal NN} \\
 \hline
 \textit{Left bolt}[mm] & 49.69  & 50.097 & 48.22 \\
 \hline
\textit{Right bolt}[mm] & 0.151  & 1.624 & 0.394 \\
 \hline
\textit{Missing bearing}[Ns/m] & 244.37 & 1.624$\times 10^2$  & 6.186$\times 10^2$ \\
 \hline
\textit{Obstacle}\{[Ns/m],[m]\} & \{7.58$\times 10^4$, 1.769\}  &\{1.297$\times 10^5$, 1.7314\} & \{1.381$\times 10^5$, 1.223\} \\
 \hline
\textit{MSE} & 0.004  &  0.005  & 0.012\\
 \hline
 \end{tabular}
 \caption{Left bolt fault mode: simultaneous parameter estimation.}
  \label{4:table}
\end{table}

\textbf{Right bolt fault}: the bolt fault mode is active with 200 mm deviation from its nominal value. Tables \ref{5:table}-\ref{7:table} present the results of the parameter estimation algorithms corresponding to the three representations, when tracking separately the fault parameters. The MSE values show that we can indeed identify the correct fault mode. Moreover,  the fault parameter values are within 6\% of the value used to generate the faulty behavior.
\begin{table}[h!]
\centering
 \begin{tabular}{|c| c|c|}
 \hline
  \textbf{Tracked fault parameters}& \textbf{Paramater value} &  \textbf{MSE} \\
 \hline
 \textit{Left bolt}[mm] & 72.75  & 1.025 \\
 \hline
\textit{Right bolt}[mm] & \cellcolor{blue!25}197.35  & 0\cellcolor{blue!25}.029 \\
 \hline
\textit{Missing bearing}[Ns/m] & 35.71 & 1.767 \\
 \hline
\textit{Obstacle}\{[Ns/m],[m]\}  & \{1.11$\times 10^5$, 0.191\}  & 1.786  \\
 \hline
 \end{tabular}
 \caption{Right bolt fault mode (acausal polynomial representation).}
  \label{5:table}
\end{table}

\begin{table}[h!]
\centering
 \begin{tabular}{|c| c|c|}
 \hline
  \textbf{Tracked fault parameters}& \textbf{Paramater value} &  \textbf{MSE} \\
 \hline
 \textit{Left bolt}[mm] & 71.84  & 1.03 \\
 \hline
\textit{Right bolt}[mm] & \cellcolor{blue!25}187.66  &\cellcolor{blue!25}0.091 \\
 \hline
\textit{Missing bearing}[Ns/m] & 49.12 & 1.818 \\
 \hline
\textit{Obstacle}\{[Ns/m],[m]\} & \{6.12$\times 10^4$, 1.04\}  & 1.855  \\
 \hline
 \end{tabular}
 \caption{Right bolt fault mode (acausal NN representation).}
  \label{6:table}
\end{table}

\begin{table}[h!]
\centering
 \begin{tabular}{|c| c|c|}
 \hline
  \textbf{Tracked fault parameters}& \textbf{Paramater value} &  \textbf{MSE} \\
 \hline
 \textit{Left bolt}[mm] & 73.97  & 1.022 \\
 \hline
\textit{Right bolt}[mm] & \cellcolor{blue!25}198.42  & \cellcolor{blue!25}0.011 \\
 \hline
\textit{Missing bearing}[Ns/m] & 20.20 & 1.792 \\
 \hline
\textit{Obstacle}\{[Ns/m],[m]\} & \{1.05$\times 10^4$, 0.977\}  & 1.792 \\
 \hline
 \end{tabular}
 \caption{Right bolt fault mode (causal NN representation).}
  \label{7:table}
\end{table}
\textbf{Bearing fault}: the bearing fault mode is active with the viscous coefficient taking the value 5000 Ns/m. Tables \ref{9:table}-\ref{11:table} present the results of the parameter estimation algorithms, when tracking separately the fault parameters. The smaller MSE values correspond to the bearing fault mode. We note the parameter estimation error variance is roughly  3\%.

\begin{table}[h!]
\centering
 \begin{tabular}{|c| c|c|}
 \hline
  \textbf{Tracked fault parameters}& \textbf{Paramater value} &  \textbf{MSE} \\
 \hline
 \textit{Left bolt}[mm] & 0.04  & 0.412 \\
 \hline
\textit{Right bolt}[mm] & 4.40  & 0.3869 \\
 \hline
\textit{Missing bearing}[Ns/m] & \cellcolor{blue!25}5060.706 & \cellcolor{blue!25}0.03 \\
 \hline
\textit{Obstacle}\{[Ns/m],[m]\} & \{3.5$\times 10^3$, 1.37\}  & 0.367  \\
 \hline
 \end{tabular}
 \caption{Bearing fault mode (acausal polynomial representation).}
  \label{9:table}
\end{table}

\begin{table}[h!]
\centering
 \begin{tabular}{|c| c|c|}
 \hline
  \textbf{Tracked fault parameters}& \textbf{Paramater value} &  \textbf{MSE} \\
 \hline
 \textit{Left bolt}[mm] & 0.06  & 0.377 \\
 \hline
\textit{Right bolt}[mm] & 16.42  & 0.365 \\
 \hline
\textit{Missing bearing}[Ns/m] & \cellcolor{blue!25}4919.18 & \cellcolor{blue!25}0.00744 \\
 \hline
\textit{Obstacle}\{[Ns/m],[m]\} & \{1.83$\times 10^5$, 1.04\}  & 0.404  \\
 \hline
 \end{tabular}
 \caption{Bearing fault mode (acausal NN representation).}
  \label{10:table}
\end{table}

\begin{table}[h!]
\centering
 \begin{tabular}{|c| c|c|}
 \hline
  \textbf{Tracked fault parameters}& \textbf{Paramater value} &  \textbf{MSE} \\
 \hline
 \textit{Left bolt}[mm] & 0.126  & 0.377 \\
 \hline
\textit{Right bolt}[mm] & 5.25  & 0.361 \\
 \hline
\textit{Missing bearing}[Ns/m] & \cellcolor{blue!25}4845.50 & \cellcolor{blue!25}0.0032 \\
 \hline
\textit{Obstacle}\{[Ns/m],[m]\} & \{1.84$\times 10^5$, 1.04\}  & 0.378  \\
 \hline
 \end{tabular}
 \caption{Bearing fault mode (causal NN representation).}
  \label{11:table}
\end{table}

\textbf{Obstacle fault}: we simulated the high fidelity model with an obstacle at 1.1 m and a viscous coefficient with the value $10^5$ Ns/m. The parameter estimation results when tracking the fault parameter separately are shown in Tables \ref{12:table}-\ref{14:table}. The smallest MSE values were obtained for the correct fault parameters. In addition, the maximum estimation error for the fault intensity and fault location parameters is 0.2\% and 0.09\% of the nominal values, respectively.
\begin{table}[h!]
\centering
 \begin{tabular}{|c| c|c|}
 \hline
  \textbf{Tracked fault parameters}& \textbf{Paramater value} &  \textbf{MSE} \\
 \hline
 \textit{Left bolt}[mm] & 3.17  & 69.618 \\
 \hline
\textit{Right bolt}[mm] & 49.57  & 69.20 \\
 \hline
\textit{Missing bearing}[Ns/m] & 5915.78 & 67.67 \\
 \hline
\textit{Obstacle}\{[Ns/m],[m]\} & \cellcolor{blue!25}\{1.01$\times 10^5$, 1.099\}  &\cellcolor{blue!25} 0.020  \\
 \hline
 \end{tabular}
 \caption{Obstacle fault mode (acausal polynomial representation).}
  \label{12:table}
\end{table}
\begin{table}[h!]
\centering
 \begin{tabular}{|c| c|c|}
 \hline
  \textbf{Tracked fault parameters}& \textbf{Paramater value} &  \textbf{MSE} \\
 \hline
 \textit{Left bolt}[mm] & 0.178  & 69.40 \\
 \hline
\textit{Right bolt}[mm] & 44.477  & 69.054 \\
 \hline
\textit{Missing bearing}[Ns/m] & 5870.32 & 67.62 \\
 \hline
\textit{Obstacle}\{[Ns/m],[m]\} & \cellcolor{blue!25}\{9.98$\times 10^4$, 1.099\}  & \cellcolor{blue!25}0.047  \\
 \hline
 \end{tabular}
 \caption{Obstacle fault mode (acausal NN representation).}
  \label{13:table}
\end{table}

\begin{table}[h!]
\centering
 \begin{tabular}{|c| c|c|}
 \hline
  \textbf{Tracked fault parameters}& \textbf{Paramater value} &  \textbf{MSE} \\
 \hline
 \textit{Left bolt}[mm] & 0.679  & 69.384 \\
 \hline
\textit{Right bolt}[mm] & 54.353  & 69.071 \\
 \hline
\textit{Missing bearing}[Ns/m] & 5867.27 & 67.560 \\
 \hline
\textit{Obstacle}\{[Ns/m],[m]\} & \cellcolor{blue!25}\{9.98$\times 10^4$, 1.099\}  & \cellcolor{blue!25}0.0122 \\
 \hline
 \end{tabular}
 \caption{Obstacle fault mode (causal NN representation).}
  \label{14:table}
\end{table}

\section{State of the art in hybrid modeling}
\label{sec_state_of_the_art}
The key idea used to enable real-time diagnosis is to build surrogate models of complex subsystems that enable fast simulations. In other words, to hybridize a system model. The surrogate models must be compatible with the system components physics-based models, and preserve, at least in part, physical interpretability. A set of recent results \cite{DBLP:journals/corr/abs-1808-03246,DBLP:journals/corr/abs-1903-11239,DBLP:journals/corr/abs-1710-04102,Kim:198748,DBLP:journals/corr/abs-1806-01242} have the  same high level objective as in our case: to build hybrid models that are more accurate and generalizable. In \cite{DBLP:journals/corr/abs-1710-04102} the authors use physics based equations composed with  data-driven models, and one of their goals is to estimate latent variables. We also estimate latent variables during the optimization process (the force and position of the rail switch are not directly observable). In our case however, we do not necessarily have an additive error term to correct for inaccurate modeling. We learn new constitutive relations, which are equations in terms of the interface variables (force, position, and velocity). There are a few other key differences such as: (1) we preserve the physical explainability by a particular choice of representation, and encourage generalizability by learning an equation rather than an input-output map (in acausal modeling we work with equations generated by physical laws and compositions between components, and not with input-output relations). \cite{DBLP:journals/corr/abs-1808-03246} propose using a NN to learn the uncertainties (errors) of a physics-based model. In our work, the goal is somewhat different. We focus on reducing complexity rather than minimizing uncertainty. In \cite{DBLP:journals/corr/abs-1903-11239} a hybrid approach is used to learn control policies for robotic throwing. Their approach uses both analytical models to provide initial estimates of control parameters, and  learned residuals on top of those estimates to compensate for unknown dynamics. They  directly learn the residuals on control parameters (i.e., action space) with deep networks, which can be interpreted as a particular way to learn a robust controller. Robust controllers are typically used to cope with uncertainties in the models. The drawback of robust controllers is that they tend to be less accurate. Our approach can be potentially used to learn controllers as well. Care must be taken when learning state dependent control maps for unstable system. If the learning algorithm is not properly initialized, the system instability can induce instability in the learning algorithm as well \cite{IM_ACC2020}. In   \cite{Kim:198748}, the authors use a hybrid modeling and learning methodology to deal with catching in-flight objects with uneven shapes. They use a generic model for the free flying object dynamics, whose vector field  is modeled using support vector regression. Typically due to noisy measurements, using acceleration samples can hinder the learning process. Under our approach, we address this challenge by actually solving the dynamical model of the flying object during the optimization projects. The ODE acts as an inherent smoother for the trajectory. In addition, we can actually deal with DAEs as well, which are more general dynamical models. \cite{DBLP:journals/corr/abs-1806-01242} introduce learnable forward and inference models, based on graph networks, as surrogate models for dynamical system. The proposed approach is the closest one in terms of its usability for model-based diagnosis since it preserves the topological representation of the system. We can as well represent the system dynamics as a graph by obtaining the block lower triangular (BLT) representation of the system. In fact, this graphical representation is automatically generated by the DAE/ODE solver as part of the simulation process. The graph-based representation describes how physical variables (e.g., forces, velocities) are connected to each other through component constitutive equations. This graph-based representation can be used for any dynamical system, and in particular for multi-body dynamics problems addressed by \cite{DBLP:journals/corr/abs-1806-01242}.

For the type of problem we are solving, there are no established benchmarks to which we can refer. It is also challenging to perform comparison of our hybrid approach with a purely data-driven model on a level footing, without imposing additional constraints (with clear justifications) on the size or complexity of the pure ML (e.g., NN) model. More specifically, our goal was to find the simplest possible data-driven model that can augment the partial physical model in such a way that some level of explainability is preserved. 
The way we make a change to the component for which we create a surrogate model is by learning new equations that connect the component variables. We have three examples of possible maps, based on polynomial and NN representations. The optimization process was based on a quasi-Newton algorithm which is a hybrid between a first order and a second order method, depending on the step-size value. One advantage is that the optimization step-size is computed automatically during the optimization process. In practice, we never have a perfect model of the physical system. In fact, we may have different models based on the application. For example, control applications might work fine with black box models, while model based diagnosis and prognostics methods require a more detailed component-based representation. As discovered in the case of the high fidelity rail switch model, accurate models come at a cost: long simulation time (9 sec). Considering that the optimization process may require many thousands of iterations, the optimization time can be very high. The scalability of the method is more related to the number of states and algebraic variables than the size of the data-driven model added to the partial physics model, since at each iteration we need to solve a differential algebraic equation (DAE).

The time complexity of the optimization algorithm is dominated by the numerical complexity of solving the ODE/DAE and is given by $O(M\times T\times n^3)$, where $M$ is the number of the optimization iterations, $T$ is the number of total time iterations of the solver, and $n$ is the number of ODE/DAE equations. The $n^3$ term comes from inverting the Jacobian of the ODE/DAE at each time instance of the ODE/DAE simulation. The values of $M$ and $T$ depend on the particular choice of optimization algorithm, on the simulation time horizon and on the tolerance parameters of the optimization algorithms and of the ODE/DAE solvers. When hybrid systems and events are involved, $T$ can get very large, since the solver requires that every time an event is triggered, the integration of the continuous-time ODE/DAE is halted at the event instant. Then the event is processed, global event iteration is performed until convergence, and finally the simulation is restarted. The optimization complexity is valid for \textit{all} least-square problems based on ODE/DAE equality constraints.

If the dynamical system can be expressed as an ODE, we can use TensorFlow or Pytorch  ODE solver capability to compute the latent variables needed for the evaluation of the loss function and its gradients. If the system admits a DAE then we can use DAE solvers that include sensitivity analysis (e.g., IDAS, CVODES). They compute the sensitivity of the latent variables with respect to the optimization variables. The training time for the rail switch example was roughly 2 hours since with the simplification we made to the rail model, the entire model can be simulated in under 0.5 sec.

\section{Conclusions}
\label{sec_conclusions}
We proposed a hybrid modeling approach to simplify a high fidelity model of a rail-switch system. In particular, we used simplified representations for the rail component using machine learning inspired models. The representations preserved the physical interpretation of the rail component. The model complexity of the model abstractions (i.e., number of equations) is reduced by two orders of magnitude. A similar reduction in the order of magnitude is obtained with respect to the simulation time of the rail switch model over a full motion cycle of the rail. The new model abstractions were used for the rail fault diagnosis. The rail switch model was augmented with additional behavior to include parameterized fault modes. An optimization based approach was used to estimate the fault parameters. We demonstrated that using  algorithms that track separately the fault parameters of each of the four fault modes produce accurate diagnosis results. The MSEs and the parameter values are used by the diagnosis engine to produce a diagnosis solution. We compared the proposed hybrid modeling approach with the state of the art.
\bibliographystyle{plain}
\bibliography{paper}

\end{document}